\documentclass[final,3p,times]{elsarticle}

\usepackage[english]{babel}
\usepackage{natbib}

\usepackage{amssymb}
\usepackage{amsmath}
 \usepackage{amsthm}
 \usepackage{caption}
\usepackage{subcaption}

\journal{The Physics of the Dark Universe}

\begin{document}

\begin{frontmatter}



\title{Gravitational Waves dynamics with Higgs portal and U(1)~x~SU(2) interactions }
\author[1]{Lucia A. Popa} 
\ead{lpopa@spacescience.ro}
\address{Institute of Space Sciences (ISS/INFLPR subsidiary),  Atomi\c{s}tilor 409, Magurele-Ilfov, Ro-077125, Romania}

\begin{abstract}

Finding the origins of the primordial Gravitational Waves (GWs) background by the near-future Cosmic Microwave Background (CMB) polarisation experiments
is expected to open a new window Beyond the Standard Model (BSM) of particle physics, allowing to investigate the possible connections between the Electroweak (EW) symmetry breaking scale and the energy scale of inflation.
 We investigate the GWs dynamics in a set-up where the inflation sector is represented by a mixture of the SM Higgs boson 
and an U(1) scalar singlet field non-minimally coupled to gravity and a spectator sector reprezented by an U(1) axion and a SU(2)  non-Abelian gauge field, 
assuming that there is no coupling, up to gravitational interactions, between inflation and spectator sectors.

We show that a mixture of Higgs boson with a heavy scalar singlet with large vacuum expectation value ({\it vev}) is a viable model of inflation 
that satisfy the existing observational data and the perturbativity constraints, avoiding in the same time the EW vacuum metastability as long as the Higgs portal interactions lead to positive tree-level threshold corrections for SM Higgs quartic coupling. \\
We evaluate the impact of the Higgs quartic coupling threshold corrections on the GW sourced tensor modes  while accounting for the consistency and backreaction constraints and show that the Higgs portal interactions enhance the GW signal sourced by the gauge field fluctuations
in the CMB B-mode ploarization power spectra. 

We address the detectability of the GW sourced by the gauge field fluctuations in presence of Higgs portal interactions 
for the experimental configuration of the future CMB polarization LiteBird space mission.
We find that the sourced GW tensor-to-scalar ratio in presence of Higgs portal interactions 
is enhanced to a level that overcomes the vacuum tensor-to-scalar ratio by a factor ${\mathcal O}(10)$, 
much above the detection threshold of the LiteBird experiment, 
in agreement with the existing observational constraints on the curvature fluctuations 
and  the allowed parameter space of Higgs portal interactions.

A large enhancement of the sourced GW can be also detected by experiments such as pulsar timing arrays and laser/atomic interferometers.
Moreover, a significant Higgs-singlet mixing can be probed at LHC by the measurement of the production cross sections for Higgs-like states, 
while a significant tree level threshold correction of the Higgs quartic coupling can be measured at colliders by ATLAS and CMS experiments.
\end{abstract}

\begin{keyword}
Cosmic Microwave Background, polarization, inflation, axions, gauge fields, vacuum stability
\end{keyword}
\end{frontmatter}

\section{Introduction}
\label{intro}

Detection of the primordial Gravitational Wave (GWs) background  would provide strong evidence for the existence 
of the cosmic inflation \cite{Guth,Linde,Albrecht}.
The properties of  scalar density fluctuations  measured by the Cosmic Microwave Background (CMB) experiments are in agreement with 
the main predictions of the simplest models of inflation 
\cite{Mukanov,Staro,Bardeen} while  the CMB polarization data place constraints on the primordial tensor modes  \cite{Planck_infl}.\\
The amplitude of the GWs background is parameterized by the tensor-to-scalar ratio $r$, the ratio of the amplitudes
of tensor and scalar power spectra of density fluctuations.
Currently, the tightest upper bound on 
tensor-to-scalar ratio i$r < 0.036$ at 95\% confidence is placed from the combined analysis of {\sc Planck} and BICEP2/Keck array 
\cite{Bicep2-Planck}.\\
The detection of the GWs background is likely  to come  from the future CMB
B-mode polarization  experiments, such as the LiteBIRD satellite experiment \cite{Lite1,Lite2} and the ground-based CMB-S4 experiment \cite{CMB-S4} 
expected to  measure the  tensor-to-scalar ratio with a sensitivity $\delta r < 10^{-3}$. 

However, the  detection of $r $ is not enough to discriminate
among the possible origins of the GWs background. \\
In the standard scenario where inflation is driven by a slow rolling scalar field minimally coupled to gravity, the GWs background 
is produced by the quantum vacuum fluctuations of the metric \cite{Guth,Gris,Staro79} and the tensor-to-scalar ratio 
can be directly related  to the Hubble expansion rate $H$ during inflation \cite{Lyth}. 
The current upper bound on tensor-to-scalar ratio
gives $H< 4.37 \times 10^{13}$ GeV.  In this scenario the GW power spectrum is nearly scale-invariant,  
nearly Gaussian and parity-conserving (non-chiral). 

This does not hold if the tensor modes are produced by the mater gauge fields during inflation.\\
One of the most studied mechanisms  leading to successfully  production of sourced GW by the matter  fields involves 
SU(2) non-Abelian massless gauge fields.
In this scenario the amplification of fluctuations in the gauge field sector, 
with subsequent enhancement of the tensor modes during inflation, is realized breaking the conformal invariance of the  gauge field
by coupling with an U(1) pseudo-scalar axion field via the Chern-Simons interaction term
  \cite{Ashead_step,Ashead_traj}. This configuration leads to an isotropic and spatially homogeneous cosmological solution 
where isotropy is protected by the non-Abelian gauge field invariance \cite{Ashead_traj}. \\
The tensor modes generated by this  mechanism can leave distinctive signatures in the primordial GWs background  compared to the
 vacuum fluctuations of the metric. 
The sourced GWs are scale-dependent \cite{Namba16}, strongly non-Gaussian 
\cite{Barnaby,Sorbo,OO,Ag1,Ag2} and parity-beaking \cite{Shi,Lue,Adshead13}. 
However, this model results in  too red spectrum of the scalar perturbations (too small   scalar spectral index) 
if the constraint on axion decay constant $f < M_{pl}$ is respected ($M_{pl}$~is the Planck mass) and is excluded 
by the observations \cite{Namba13,Pajer,Emanuela13}.

Accommodation of  the observational bounds for the scalar and tensor   modes while
the gauge fields play an important role in production of cosmological perturbations is 
challenging. During inflation a large amount of spin-2 particles  produced  to source the GWs background
at observational level causes backreactions on dynamics of the axion-gauge fields \cite{Emanuela19,Oxana,Komatsu_new,Komatsu_ref} that can result 
in predictions for cosmological observables ruled out by the experimental measurements.
The solution  was to consider the axion-gauge field sector as spectator,  decoupled up to
gravitational interactions from the inflation sector  \cite{Namba16}, set-up used by most of the sourced GWs studies in context of the standard 
inflation \cite{Emanuela19, Komatsu_new,Komatsu_ref,Thorne}.

Finding the GW origin by using the CMB polarisation measurements is expected to open a new window beyond the Standard Model (BSM) 
of particle physics,
allowing to investigate the possible connections between the energy scale of inflation and the subsequent energy scales in the evolution of the Universe.

The discovery by the Large Hadron Collider (LHC) of the Higgs boson \cite{ATLAS,CMS} increased the interest in
so called Higgs portal interactions that connect the hidden (dark)  and visible sectors
of the Standard Model (SM).
Scenarios beyond-the-SM (BSM), that introduce a dark sector in addition to the visible SM sector are
required to explain a number of observed phenomena in particle physics, astrophysics and
cosmology such as non-zero neutrino masses and oscillations, nature of Dark Matter (DM),
 baryon asymmetry of the Universe,  cosmological inflation \cite{Beachaman}. \\
In the original  Higgs inflation model \cite{Bezrukov1,Bezrukov2}  the SM
Higgs boson can play the role of inflaton  if it has  a large non-minimal coupling to gravity  $\xi \sim {\cal O} (10^{5}) $.
The model predictions are  consistent with cosmological data \cite{Planck_infl},
suggesting  possible connections between electroweak (EW) scale and the inflation scale. 
However, for such large values of $\xi$ the unitary bound scale $\Lambda_{U} \sim  M_{pl} / \xi$ could be close or much below the energy 
scale of inflation. Such large couplings can be generated  at Renormalisation Group (RG) loop levels, 
but in this case  the price to pay is the vacuum metastability, as
the SM Higgs quartic coupling becomes negative due to the radiative corrections \cite{Bezrukov3,Degrassi}. 

Higgs portal interactions allow  to build viable inflation models where the 
inflaton  role is played by a mixture  of  Higgs boson with a  singlet scalar field 
with non-zero vacuum expectation value ({\it vev})  non-minimally coupled to gravity \cite{Lebedev_stability,Lebedev2,higgs_portal}.
Such models are in agreement with the CMB observations \cite{Planck_infl}
and satisfy the perturbativity and the tree-level unitarity constraints as long as the effective mixing coupling $\lambda_{hs}$
is sufficiently small.
The Higgs-scalar singlet  mixture  leads to  positive tree-level threshold 
correction to the Higgs quartic coupling, preventing in this way the EW vacuum metastability \cite{Lebedev_stability,Ellis,Lebedev2}.

Higgs-scalar singlet models are also attractive from the standpoint of the GW production 
as they predict strong first-order cosmological phase transitions
(PTs) at the EW scale, leading  to detectable GW signal. 
The PT can be generated  when the scalar field experiences a change in its {\it vev}  
in a certain temperature range, undertaking a transition into a new phase by quantum or thermal processes.
This process proceed by the nucleation of bubbles in plasma through:
(b) collisions of expanding bubble walls, (s) sound waves produced by 
the bulk motion in plasma after the bubbles have collided but before expansion has
dissipated the kinetic energy  and  (t) magnetohydrodynamic turbulence in  plasma after the bubbles have
collided (for details see  Ref. \cite{Chiara_Caprini} and references therein).
As these processes coexist,   the present GW energy density spectrum can be approximated as a linear combination 
of the contributions from the above mechanisms: $\Omega_{GW}$=$\Omega_{b}$+$\Omega_{s}$+$\Omega_t$.  \\
The peak-integrated sensitivity curves (PISCs) for $\Omega_{b}$, $\Omega_{s}$ and $\Omega_t$    
obtained in the Higgs-scalar singlet model, corresponding  to different future GW searches,
are given in Refs. \cite{Fresh,Fresh1}. \\
The analysis of the effects of the Higgs-scalar singlet couplings on the EW phase transition,
including the dimension-six non-renormalizable operators 
to couple the singlet scalar field with the SM Higgs doublet \cite{Economu}, shows  that
the EW phase transition can occur in two-steps, a singlet
phase transition at high temperature and a subsequent strong first-order phase transition in SM sector.
This scenario can explain the baryon asymmetry in the 
Universe and is compatible with scalar singlet as Dark Matter candidate with a mass nearly half of the Higgs mass.\\
The consequences of the EW symmetry breaking on GW energy spectrum in presence of axion-Higgs  
non-perturbative couplings  were recently analyzed in context of $R^2$ inflation and Einstein-Gauss-Bonnet inflation models
\cite{Economu1,Economu2}.

A significant Higgs-singlet mixing can be probed at LHC 
by  measurement of the production cross sections for Higgs-like states \cite{Falk,PDG}.
Furthermore, the Higgs-singlet mixing can lead to a significant tree level modification of the Higgs quartic coupling, which can
be measured at colliders \cite{ATLAS,CMS}.

In this paper we investigate a scenario where the axion and non-Abelian gauge fields are confined to the spectator sector while the inflation sector is represented by a mixture of Higgs and a scalar singlet field non-minimally coupled to gravity. 
We show that the Higgs portal interactions lead to  positive tree-level threshold corrections to SM Higgs quartic coupling leading to the change of the Hubble expansion rate during inlation. This impacts on the evolution of 
the axion-gauge field spectator sector modifying 
 the time-dependent mass parameter of the gauge field fluctuation.\\
We assume that there is no coupling, up to gravitational interactions, between inflation and spectator sectors and 
the background energy density is dominated by  inflation. 
The relevant Jordan frame action of the model, where hat is representing the quantities in the Jordan frame, is given below:
\begin{eqnarray}
\label{action}
{\cal S}_J &  = &  \int  {\rm d}^4 x \sqrt {-{\hat g}} \,  [  \frac{1}{2} (M^2_{pl}+\xi_hh^2+\xi_{s} s^2) {\cal {\hat R}} \nonumber \\ 
& - & \underbrace{ \frac{1}{2}(\partial_{\mu}h)^2  
   - \frac{1}{2}( \partial_{\mu} s)^2 +V(h,s)}_{ {\cal L}_{inf} } 
 -  \underbrace{\frac{1}{2}  (\partial \hat{\chi})^2   -V_a(\hat {\chi})  -\frac{1}{4}\hat{F}^a_{\mu\nu} \hat{F}^{a\mu\nu}+ {\hat{\cal L}}_{int} }_{ {\cal L}_{spec}} ]  \,.
\end{eqnarray}
Here $\hat{g}$ is the determinant of the metric tensor, $M_{pl}$ is the Planck mass, ${\cal {\hat R}}$ is the Ricci curvature, $\xi_{h}$ and 
$\xi_s$ are the couplings of Higgs 
and scalar fields with  the  curvature,  $V(h,s)$ is the inflaton potential,  $V_a(\hat {\chi} )$ is the axion potential and 
$\hat{F}^a_{\mu\nu}$ is the gauge strength field tensor. The axion field  is expected to interact with the gauge fields 
through the Chern-Simons interaction term $\hat {\cal L}_{int}$. Transition of the action (\ref{action}) from the Jordan to the Einstein frame  
is accomplished by rescaling the metric:
\begin{eqnarray}
\label{conform}
g_{\mu \nu} = \Omega {\hat g}_{\mu \nu}\,, \hspace{0.6cm} \Omega(h,s)=(M^2_{pl}+\xi_h h^2 +\xi_{s} s^2)  \,.
\end{eqnarray}

We study the dynamics of the axion-gauge field spectator in presence of Higgs portal interactions in a comprehensive parameter space 
for two  slow-roll solutions for the mass parameter of gauge field fluctuations \cite{Komatsu_new} while accounting for 
consistency and backreaction constraints. \\
The most interesting result obtained is the enhanced production of the sourced GW by the non-Abelian gauge field in presence of Higgs portal interactions 
to a level that overcomes the quantum vacuum fluctuations by a factor ${\mathcal O}$(10) for both solutions, 
much above the detection threshold of the near-future B-modes polarization experiments, in agreement with the CMB observations 
on curvature fluctuations and with the allowed parameter space of Higgs portal interactions. 
 
The paper is organized as follows.  In section~\ref{inflation} we review the Higgs-singlet inflation model and place constraints 
on Higgs portal parameter space requiring the agreement with the observations on curvature fluctuations.
In section~\ref{CNI}  we present the Higgs-singlet inflation model with transiently rolling U(1) x SU(2) spectator fields and analyse
the consistency and backreaction constraints. Section~\ref{GW} 
is dedicated to provide constraints on the parameter space of the spectator axion-gauge field model in presence of Higgs portal interactions.
We present our conclusions in section \ref{final}. 

Throughout the paper we consider an homogeneous and isotropic 
background described by the Friedmann-Robertson-Walker (FRW) metric and work  in natural units ($\hbar$=$c$=$M_{pl}$=1) 
unless specified otherwise.

\section{Higgs-scalar singlet inflationary dynamics}
\label{inflation}
Extension of the Higgs sector with a real scalar field in presence of the large couplings to scalar curvature 
can lead to inflation based on scale invariance of the Einstein frame 
scalar potential at large field values. As in the case of Higgs inflation  \cite{Bezrukov1,Bezrukov2}
the potential becomes exponentially flat at  large field values that is favoured by the inflationary observables \cite{Planck_infl}. Below we summaise the basic ideas of the Higgs-scalar singlet inflation, following Refs. \cite{Lebedev_stability,Lebedev2}. 

We assume a ${\mathbb Z}_2$-symmetric inflation potential of the form:
\begin{eqnarray} 
\label{Vhs}
V(h,s)=\frac{1}{4} \lambda_h h^4 +\frac{1}{4} \lambda_{hs} h^2 s^2 +
 \frac{1}{4} \lambda_{s} s^4 + \frac{1}{2}m^2_h h^2 +\frac{1}{2}m^2_{s} s^2 \,,
\end{eqnarray}
where $m_h$ and $m_s$ are $h$ and $s$ fields masses, $\lambda_h$ and $\lambda_s$ are their quartic couplings and
 $\lambda_{hs}$  is the mixing quartic coupling.
Here after  we will consider that  both fields develop non-zero vacuum expectation values
 (${\it vev}$) denoted by $<h> = {\it v}$, $< s > = {\it w}$ and $w \gg { \it  v}$.  
We will also consider large field values and $\xi_h \gg \xi_s$ such that:
\begin{equation}
\label{large_field}
\xi_h h^2+\xi_{s} s^2 \gg 1\,,\hspace{1cm} \xi_h +\xi_{s} \gg 1\,.
\end{equation}
After the conformal transformation given in Eq. (\ref{conform}), the  kinetic term and the 
inflation potential in the Einstein frame read as \cite{Lebedev_stability}:
\begin{eqnarray}
\label{lkin}
{\cal L}_{kin} & = & \frac{1}{2} (\partial_{\mu} \phi)^2 + \frac{1}{2} \frac{\xi^2_h \tau^2 + \xi_{s}^2}
{(\xi_h \tau^2 +\xi_{s})^3} (\partial_{\mu}\tau)^2 \,,\\
\label{U}
U (\tau)& = &\frac{ \lambda_h \tau^4+\lambda_{h s} \tau^2 + \lambda_{s}}{4(\xi_h\tau^2+\xi_{s})^2} \,,
\end{eqnarray}
where  the new variables $\phi$ and $\tau$ are defined as:
\begin{eqnarray}
\label{phi}
\phi =  \sqrt{ \frac{3}{2}} \log{ (\xi_h h^2 +\xi_{s} s^2) }\,, \hspace{1cm}
\tau  =  \frac{h}{s} \,.
\end{eqnarray}

The minima of the potential  given in Eq. (\ref{U}) are classified in Refs. \cite{Lebedev_stability,Lebedev2} according to  the particle content during inflation.
For Higgs-singlet inflation to occur  the following conditions are required:
\begin{eqnarray}
\label{HS_exist}
2 \lambda_h\xi_{s} -  \lambda_{h s} \xi_h > 0\,,\,\,\, 
2\lambda_{s}\xi_h-\lambda_{h s}\xi_{s}> 0\,, \,\,\, 
\tau_{min}= \sqrt{ \frac{2 \lambda_{s} \xi_h -\lambda_{h s} \xi_{s}} 
{ 2 \lambda_h \xi_s - \lambda_{h s} \xi_h} } \,,
\end{eqnarray}
As $\tau \sim M_{pl}/ \sqrt{\xi_h} $  it can be integrated out in  Eq. (\ref{lkin}), leaving $\phi$ the only dynamical variable during inflation. 
Details of this computation can be found  in Refs. \cite{Lebedev_stability,Lebedev2,Falk,Scalaron}.

The potential for $\phi$ can be written as:
\begin{equation}
\label{U_phi}
U(\phi)=\frac{\lambda_{eff}}{4 \xi^2_h } 
\left( 1+ \exp{ \left(-\frac{2 \phi}{\sqrt{6}}\right) }\right)^{-2} \,,
\end{equation}
where $\lambda_{eff}$ is given by:
\begin{eqnarray}
\lambda_{eff}=\frac{1}{4} \frac{4 \lambda_{s}\lambda_h-\lambda^2_{hs}}
{\lambda_{s}+\lambda_h x^2 -\lambda_{h s}x}\,,
\hspace{0.5cm} x=\frac{\xi_s}{\xi_h} \,.
\end{eqnarray}
The potential from Eq. (\ref{U_phi}) leads to the following slow-roll parameters:
\begin{eqnarray}
\label{epsilon_phi}
\epsilon_{\phi} &=& \frac{1}{2} \left( \frac{U_{\phi}(\phi)}{ U(\phi) }\right)^2 \simeq \frac{4}{3}\exp{\left(-\frac{4 \phi}{ \sqrt{6}}\right)} \,, \\
\label{eta_phi}
\eta _{\phi}&=& \frac{U_{\phi,\phi}(\phi) }{U(\phi)} \simeq-\frac{4}{3}\exp{\left(-\frac{2\phi}{\sqrt{6}}\right)} \,,
\end{eqnarray}
where we denote $U_{\phi}(\phi)\equiv \partial U(\phi)/\partial \phi$ and $U_{\phi,\phi}(\phi)\equiv \partial U_{\phi}(\phi)/\partial \phi$. \\
During inflation $e^{\phi} \gg 1$ and $\epsilon_{\phi}\,, \eta_{\phi} \ll1$. Inflation ends when $\epsilon_{\phi} \simeq1$ corresponding to:
\begin{equation}
\phi_{end} \simeq \sqrt{ \frac{3}{4} } \ln{\frac{4}{3}}
\end{equation}
The number of e-folds before the end of inflation can be obtained as:
\begin{eqnarray}
N= - \int^{\phi_{end}}_{\phi_{in}} \frac{U(\phi)}{U_{\phi}(\phi)}\,\, {\rm d} \phi \simeq 
                                   \frac{3}{4} \exp{( 2 \phi_{in}/ \sqrt{6})} \,,
\end{eqnarray}
leading to the value of the field at beginning of inflation:
\begin{equation}
\phi_{in} \simeq  \frac{\sqrt{6}}{2} \ln{ \frac{4N}{3} }\,.
\end{equation}

For $N=59$ e-folds, as required  by the {\sc Planck} normalization \cite{Planck_cosmo},  
$\phi_{in}=5.34M_{pl}$, while  the value of the inflaton field $\phi_*$ corresponding to the Hubble crossing of the largest observable CMB scale  at $N \simeq 55$ e-folds before the end of inflation \cite{Planck_infl} is $\phi_*=5.26 M_{pl}$.

The  power spectra of curvature  and  tensor perturbations 
generated during inflation by the quantum vacuum fluctuations are: 
\begin{eqnarray}
\label{vac_ps}
{\cal P}^{v}_{{\cal \zeta}} (k) & = &  A_s\left (\frac{k}{k_0}\right)^{n_s-1}\,, 
\hspace{0.5cm}A_s=  \frac{H^2}{24\pi^2 \epsilon_{\phi}}\,, \\
{\cal P}^{v}_t (k) & = & A_t \left (\frac{k}{k_0}\right)^{n_t}\,,
\hspace{1cm} 
A_t  =  \frac {2 H^2 }{ \pi^2}\,, 
\end{eqnarray}
where  $A_s$ and $A_t$  are the scalar and tensor power spectra amplitudes, $n_s$ and $n_t$  the corresponding spectral indexes 
and $k_0$ is the pivot scale.
The vacuum tensor-to-scalar ratio at $k_0$ is defined as $r^{(v)}_{k_0}={\cal P}^{(v)}_t /P^{(v)}_{\cal R}$. 

To the first order in slow-roll approximation, $n_s$, $n_t$ and $r^{vac}$ are:
\begin{eqnarray}
n_s \simeq 1- 6 \epsilon_{\phi}+2 \eta_{\phi} \,, 
\hspace{0.5cm}n_t=-2 \epsilon_{\phi}\,,\hspace{0.5cm}r^{vac}=16\epsilon_{\phi}\,,
\end{eqnarray}
leading to $n_s=0.961$ and $r^{(v)}=3.44 \times 10^{-3}$ for $\phi_*=5.26 M_{pl}$.
The best fit of {\sc Planck} measurements in the standard $\Lambda CDM$ cosmology indicates 
$ {\cal P}^{obs}_{\zeta} =(2.1 \pm 0.03 ) \times 10^{9} $ and $n_s=0.9649 \pm 0.0042$  (65\% CL) 
at  $k_0=0.05$Mpc$^{-1}$ 
\cite{Planck_infl,Planck_cosmo} while the joint analysis of BICEP2/KECK and Planck data constrained 
the vacuum tensor-to-scalar ratio $r^{(v)}_{0.05} < 0.036$ ( 95\% CL)  \cite{Bicep2-Planck}. 

\subsection{Higgs portal assisted Higgs-singlet inflation}

Higgs portal interaction term $V(h,s) \subset \lambda_{h,\phi} h^2 s^2$ from Eq. (\ref{Vhs})
has  distinct contributions  to both  EW scale 
and  to the high energy  scales,
ensuring the the stability of the inflation potential.  \\
In the limit $\lambda_s{\it  w^2} \gg {\lambda_h \it v^2}$, the extremization of the scalar potential $V(h,s)$ leads to 
the squared Higgs mass eigenvalue~\cite{Lebedev_stability,Ellis}:
\begin{eqnarray}
\label{mass_eign}
m^2_h  \simeq  2 {\it v^2} \left[ \lambda_h -\frac{\lambda^2_{hs} }{4 \lambda_s}    \right]  \,,
\end{eqnarray}
$m_s^2=2 \lambda_s {\it w}^2 +2(\lambda^2_{hs}/\lambda_s){\it v}^2 $ and mixing angle 
$\tan 2 \theta=\lambda_{hs}{\it v}{\it w}/ (\lambda_h{\it v}^2 -\lambda_s {\it w}^2)$.

The Higgs ${\it vev}$ is fixed at 
${\it v} \equiv (\sqrt {2} G_F)^{1/2}$= 246.22 GeV by the Fermi constant $G_F$ while 
the measured SM Higgs mass is  $m^{SM}_h =125.10 $ GeV \cite{PDG} leading to  
$\lambda^{SM}_h  \simeq 0.128 $. 
As a result,  Eq. (\ref{mass_eign}) shows that the same value of $m^{SM}_h$ can be obtained 
for  various values of $\lambda_h$ as long as  $ \lambda_h - \lambda^2_{hs}/ \ 4 \lambda_s  \simeq \lambda^{SM}_h$.\\
Consequently, at the EW scale the quartic  couplings $\lambda_i $ are all positive 
while the perturbativity constraint ($\lambda^2_i / 4 \pi<1$) imposes $\lambda_i < 1$.
Therefore, in the Higgs-singlet inflation  the Higgs quartic coupling is  given by:
 \begin{eqnarray}
 \label{prag}
 \lambda_h \bigg |_{Higgs+singlet} = \lambda^{SM}_h+ \frac{\lambda^2_{hs}}{4 \lambda_s} \,.
 \end{eqnarray}
This threshold effect occurs at tree level, avoiding the instability of the EW vacuum 
and is dominant over quantum loop contributions. Moreover, the size of the shift 
$\lambda_{hs}^2/4 \lambda_s$ does not depend on the singlet mass, allowing to prevent the potential instability
at large field values \cite{Ellis}.

On the other hand, the couplings that appear in the inflation potential, including the non-minimal couplings $\xi_h$ and $\xi_s$,
are associated to different energy scales encoded in the renormalisation group (RG) equations \cite{Ema,higgs_portal,Donald}.
Connecting small and large scales in the large field limit $\xi_h \phi^2 +\xi_s s^2 \gg1$ can be challenging as
the scalar propagators are modified by the curvature term and RG equations receive corrections from higher dimensional operators  
that can not be calculated reliably \cite{Lebedev_stability,Lebedev2}.

\begin{figure}
\begin{subfigure}[b]{0.3\textwidth}
\centering

\vspace{-3cm}
\includegraphics[width=9cm,height=5.cm]{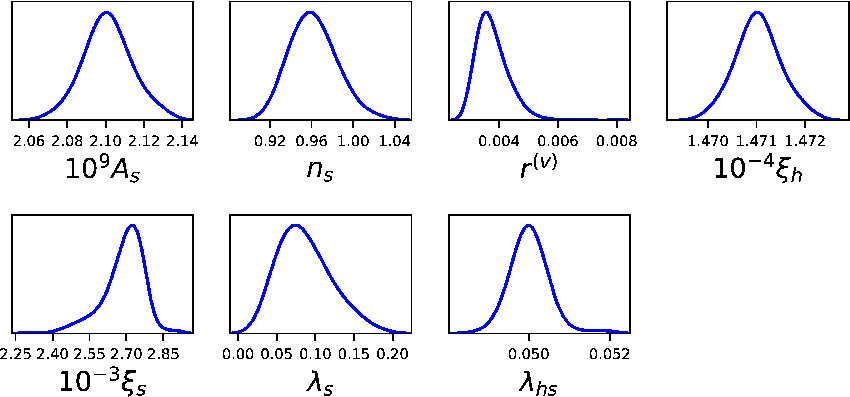}
\end{subfigure}
\begin{subfigure}[b]{1\textwidth}
  \centering
  \includegraphics[width=4.5cm,height=4.5cm]{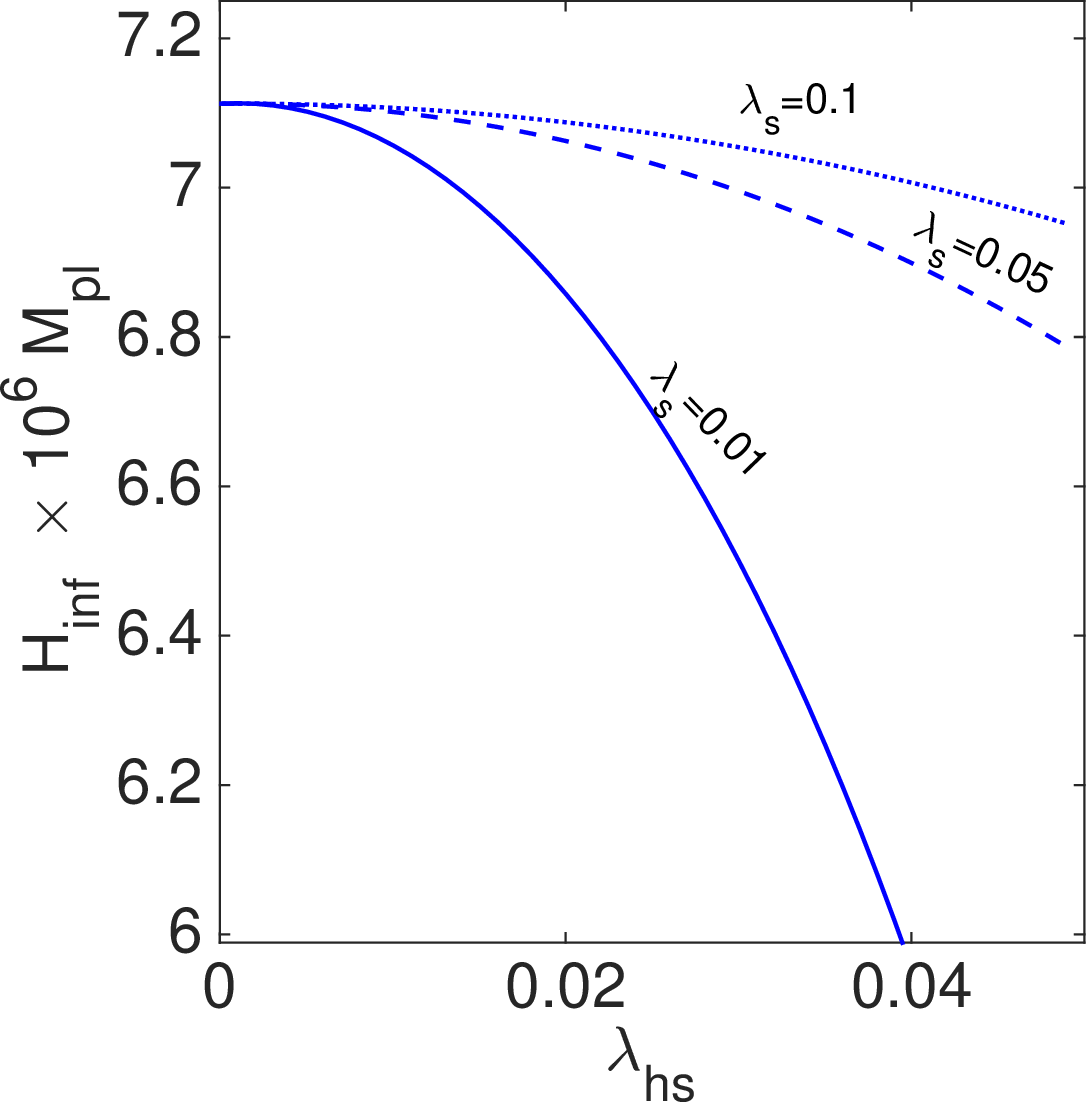}
 \end{subfigure}
\caption{{\it Left}: The marginalised likelihood probability distributions obtained for Higgs-singlet inflation model parameters.
{\it Right}:  Evolution of the Hubble expansion rate during inflation $H_{inf}$ with  $\lambda_{hs}$ for different values of $\lambda_s$. 
Here $H_{inf}$ is evaluated at $\phi_* =5.26\,M_{pl}$ corresponding to the Hubble crossing 
of the largest observable CMB scale at $N\simeq 55 $ e-folds before the end of inflation \cite{Planck_infl}.\label{Fig1}}
\end{figure}

In what follows we do not impose quantum corrections to the Higgs-singlet model. Instead we evaluate
$\lambda_h$ at EW scale considering the threshold effect given by Eq. (\ref{prag}). \\
We adopt the {\sc Planck} normalisation ${\cal P}^{obs}_{\zeta}$ at $k_0=0.05$~Mpc$^{-1}$ to fix the ratio
$U(\phi)/\epsilon_{\phi}=24 \pi^2 {\cal P}^{obs}_{\zeta}$ \cite{Bezrukov1} 
and evaluate the parameter space allowed by Higgs-singlet  model  taking as target model 
the standard $\Lambda CDM$ model with the best fit parameters obtained by {\sc Planck} \cite{Planck_cosmo}.\\
Our numerical analysis is done as follows. 
We choose $\lambda_s$ and $\lambda_{hs}$  as free parameters in the range $0<\lambda_i<1$ and take $\xi_h$ as free parameter while 
$\xi_s$ is constraint by  $\lambda_{hs} \xi_h/2 \lambda_h<\xi_s < 2\lambda_{s}\xi_h / \lambda_{hs}$ as required by Eq. (\ref{HS_exist}).\\
We modify the original CAMB code\footnote{http://camb.info} \cite{camb}
to numerically compute the slow-roll parameters and inflationary observables $A_s, n_s, r_{v}$ for Higgs-singlet model at $k_0=0.05 Mpc^{-1}$
and use the Monte-Carlo Markov Chains technique\footnote{http://cosmologist.info/cosmomc/} 
to sample from the space of  Higgs-singlet inflation model 
parameters and generate estimates of their posterior distributions. The tensor spectral index $n^{v}_t$ is obtained 
from the consistency relation $n^{v}_t=-r^{v}/8$.
We assume a flat universe and uniform priors for all free parameters adopted in the analysis.\\
Left panel from Figure \ref{Fig1} presents the posterior likelihood probability distributions for the Higgs-singlet model parameters.
Requirement that {\sc Planck} normalisation should be satisfied  results in tight constraints for all parameters.\\
The confidence intervals (at 99\% CL) of parameters that we will use in this  analysis are given below:
\begin{eqnarray}
\label{confidence}
\lambda_{hs}& : &\hspace{0.5cm}4.7\times 10^{-2} \,\,\, \div \,\,\,5.2 \times 10^{-2}   \\
\lambda_s &:& \hspace{0.5cm}   1.5\times 10^{-2}\,\,\, \div \,\,\,1.9 \times 10^{-2} \nonumber \\
\xi_h & : & \hspace{0.5cm}1.469 \times 10^{4} \,\,\, \div \,\,\, 1.473 \times 10^{4}\nonumber\\
\xi_s &:&  \hspace{0.5cm} 2.39 \times 10^{3} \,\,\, \div \,\,\, 2.83 \times 10^{3} \ \nonumber\\
H_{inf} & : &\hspace{0.5cm} 2.94 \times 10^{-6}M_{pl}\,\, \div \,\,7.02 \times 10^{-6} M_{pl}\nonumber   
\end{eqnarray}
Here $H_{inf}$ is a derived parameter evaluated at $\phi_{*} =5.26\,M_{pl}$
corresponding to the Hubble crossing of the largest observable CMB scale at $N\simeq 55 $ e-folds before the end of inflation \cite{Planck_infl}.
The right panel from Figure \ref{Fig1} shows the evolution of the Hubble expansion rate during inflation $H_{inf}$ with  $\lambda_{hs}$ for increasing values of $\lambda_s$. 
The figure shows that  $H_{inf}$  increases when  
the tree-level threshold corrections to SM Higgs quartic coupling are decreased.\\
We show that a mixture of Higgs boson with a heavy scalar singlet with large {\it vev}
is a viable model of inflation that satisfy the {\sc Planck} data constraints avoiding 
at the same time  the instability of the EW vacuum as log as the Higgs portal interactions
lead to a positive tree-level threshold corrections for SM Higgs quartic coupling. 
Moreover, these corrections lead to changes of the Hubble expansion  during inlation that impact on the evolution of the axion-gauge field spectator sector.

\begin{figure}
\centering
\includegraphics[width=12cm,height=6.cm]{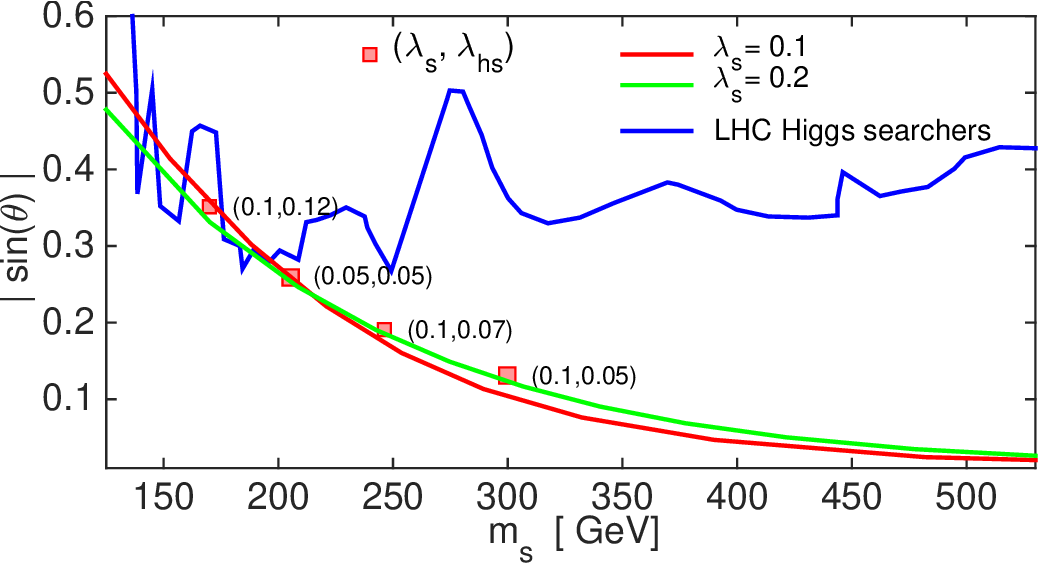}
\caption{Maximal allowed values for $|\sin(\theta)|$ in the scalar-singlet high mass region $m_s~\in~[125 - 600]$ GeV
from the direct LHC Higgs searchers \cite{Robens} (blue) compared with 
$m_s$ and $|\sin(\theta)|$ values obtained in Higgs-scalar singlet model for $\lambda_s$=0.1 (red) and $\lambda_s$=0.2 (green) when  $\lambda_{hs}$ is allowed to vary in the confidence interval given in Eq. (\ref{confidence}). Some particular 
values  $(\lambda_s,\lambda_{hs})$ are also indicated. The figure shows that
the Higgs-singlet mixing can lead to a significant tree-level modification of the Higgs quartic coupling which can
be measured at colliders \cite{ATLAS,CMS}.\label{Fig2} }
\end{figure}

We evaluate the scalar-singlet mass $m_s$ and mixing angle $|\sin(\theta)|$ for 
$\lambda_s$ and $\lambda_{hs}$ in the confidence intervals given in Eq. (\ref{confidence}). 
The best fit vales $(\lambda_s,\lambda_{hs})=(0.1, 0.05)$ lead to $m_s=289.42$ GeV and $|\sin(\theta)|=0.122$. \\
One should note that for $m_s \simeq 290$ GeV the maximal allowed mixing angle is 
$|\sin(\theta)|_{max}=0.31$ and the maximal and minimal allowed branching ratios  are 
$BR^{H\rightarrow hh}_{max}=0.4$ and $BR^{H\rightarrow hh}_{min}=0.18$ at 95\% CL \cite{PDG} 
while the upper bound of the invisible Higgs boson branching ratio is 
$BR^{H\rightarrow hh}_{inv} < 0.11$ at 95\% CL \cite{Economu}. 
Figure \ref{Fig2} presents $m_s$ - $|\sin(\theta)|$ 
dependences obtained in Higgs-scalar singlet model for $\lambda_s$=0.1  and $\lambda_s$=0.2 when  $\lambda_{hs}$ is allowed to vary, compared with 
the maximal allowed values for $|\sin(\theta)|$ in the scalar-singlet high mass region $m_s \in [125 - 600]$ GeV
from direct LHC Higgs searchers \cite{Robens}. 
The figure clearly shows that
the Higgs-singlet mixing can lead to a significant tree-level modification of the Higgs quartic coupling which can
be measured at colliders \cite{ATLAS,CMS}.

\section{Higgs-singlet inflation with transiently rolling U(1) x SU(2) spectator fields}
\label{CNI}

Throughout we will assume that the total energy density is dominated by the energy 
density of the inflaton field and treat the Hubble expansion rate $H$ as constant during inflation.
More specifically we will take the value of $H$ at $\phi_*=5.26M_{pl}$ corresponding to the Hubble crossing of the largest observable CMB 
scale at $N\simeq 55$ e-folds before the end of inflation.

We briefly review the main features of Higgs-singlet inflation in presence of the axion-gauge fields spectator  \cite{Adshead12,Adshead13,Adshead16}. 
The Einstein frame Lagrangian of the model is given by:
\begin{equation}
\label{lagrangian}
{\cal L}=\frac{1}{2} {\cal R} - \frac{1}{2} (\partial \phi)^2 -U(\phi)-
\frac{1}{2} (\partial \chi)^2 - V(\chi) -\frac{1}{4}F_{\mu \nu} \tilde{F}^{\mu \nu} +{\cal L}_{int} \,,
\end{equation}
where   $\phi$ is the inflaton field and $U(\phi)$ is its potential as given in Eq. (\ref{phi}) and Eq. (\ref{U_phi}),
$\chi$ is the axion field endowed with shift symmetry guaranteed by the U(1) gauge invariance, 
$V(\chi)$ is the axion potential, $F_{\mu \nu}=\partial_{\mu}A_{\nu}-\partial_{\nu}A_{\mu}-g \epsilon^{abc} A^b_{\mu}A^c_{\nu}$ 
is the strength  of a non-abelian SU(2) gauge field  and 
${\tilde F}^{\mu\nu} \equiv \eta^{\mu \nu \rho \upsilon} F_{\rho \upsilon}/(2 \sqrt{g})$ is its dual, 
where $\eta^{\mu \nu \rho \sigma}$ is an antisymmetric tensor satisfying $\eta^{0123}=g^{-1}$ and $g$ is the gauge coupling. \\
The axion field is expected to interact with the gauge field through a Chern-Simons  term of the form:
\begin{eqnarray}
\label{CS}
L_{int}=- \lambda \frac{\chi}{f} F_{\mu \nu} \tilde{F}^{\nu \mu} \,,
\end{eqnarray} 
where $\lambda$ is a dimensionless coupling constant and $f$ is the axion decay constant of mass dimension.\\
To minimize the influence of spectator sector on the curvature perturbations generated during inflation and to render 
the sourced GW produced by the gauge-fields viable
we will consider a model that can lead to localized gauge field production where the spectator axion transiently rolls on potential of the form
 \cite{Barnaby12,Namba16,Ozsoy}:
\begin{equation}
\label{V_axion}
V(\chi)= \mu^4 \left[1+\cos{\left (\frac{\chi}{f} \right) }\right] \,,
\end{equation}
where $\mu$ is its modulation amplitude with mass dimension. 
The axion field rolls between $\chi_{min}$=0 and  $\chi_{max}=\pi f$ with a velocity that obtains the maximum value at $t_*$ when 
$\chi_*=0.5 \pi  f$  and the slope of the axion potential $V_{\chi}(\chi_*)$ is maximal. Here $V_{\chi} \equiv \partial_{\chi}V/\partial \chi$.

We assume an initial gauge field configuration described by \cite{Adshead12,Malek_a,Malek_b}:
\begin{eqnarray}
\label{gauge_conf}
A^a_0=0\,, \hspace{0.6cm} A^a_i  = \delta^a_ia(t)Q(t)\,,
\end{eqnarray}
where $a(t)$ is the scale factor and $Q(t)$ is the SU(2) gauge field.   This configuration leads to an isotropic 
and spatially homogeneous cosmological solution where isotropy is protected
by the non-Abelian gauge field invariance \cite{Ashead_traj}. 

The time-dependent components of the axion-gauge field  model $X=( \chi, Q, f)$ 
translate from Jordan to Einstein frame under the conformal transformation given by Eq.(\ref{conform}) as 
 $\partial X / \partial {\hat X} = \Omega^{-1/2}$ where  $ \hat{X} $ denote the Jordan frame counterpart \cite{Kaiser}. \\
In the large-field approximation given by Eq (\ref{large_field}) this leads to 
 $\Omega^{-1/2} \simeq \exp {\left(- 2 \phi_*/ \sqrt{6}\right)} \simeq 1.36 \times 10^{-2}$,
 leaving the evolution equations of the axion-gauge field spectator unchanged.
 
The evolution equations of the Hubble parameter reads as:
\begin{eqnarray}
\label{Hubble}
3 H^2 & = & \frac{1}{2} {\dot \chi}^2 +V(\chi)
 + \frac{1}{2} {\dot \phi}^2 +U(\phi) +
\frac{3}{2} \left[ ( {\dot Q} +H Q)^2 +g^2 Q^4 \right ]\,,
\end{eqnarray}
and the equations of motion for  inflaton,  axion, and gauge fields without beackreaction are given by \cite{Emanuela19,Oxana,Komatsu_new}:
\begin{eqnarray}
\label{EM_inflaton}
{\ddot \phi }+3 H {\dot \phi} +U_{\phi}(\phi) & = & 0 \,,\\
\label{EM_axion}
\ddot{\chi} + 3 H \dot {\chi} + V_{\chi}(\chi) &=& -\frac{3g\lambda}{f} Q^2 ( \dot{Q} +H Q) \,, \\
\label{EM_gauge}
\ddot{Q}+3H \dot {Q} +( \dot{H} +2 H^2)Q + 2 g^2Q^3 & = & g \frac {\lambda} {f}  \dot {\chi} Q^2\,.
\end{eqnarray}
The Hubble slow-roll parameter $\epsilon_H$
contains contributions from  inflaton, axion and gauge fields:
\begin{eqnarray}
\label{epsH} 
\epsilon_H= \epsilon_{\phi}+\epsilon_{\chi}+\epsilon_{Q_E} +\epsilon_{Q_B}\,,
\end{eqnarray}
where the corresponding slow-roll parameters:
\begin{eqnarray}
\label{eps}
\epsilon_{\phi}= \frac{{\dot \phi}^2}{2H^2}\,,\hspace{0.3cm}
\epsilon_{\chi}= \frac{{\dot \chi}^2}{2H^2}\,,\hspace{0.3cm}
\epsilon_{Q_E}= \frac{ ({\dot Q} + HQ)^2}{H^2}\,,\hspace{0.3cm}
\epsilon_{Q_B}= \frac{ g^2Q^4}{H^2}\,,
\end{eqnarray}
are assumed  to be smaller than unity during inflation. These parameters modify $\epsilon_H$ in Eq. (\ref{epsH}), that in turn affects the spectral index of scalar perturbations \cite{Emanuela19,Komatsu_ref,Fujita}:
\begin{eqnarray}
\label{ns}
n_s-1=2(\eta_{\phi}-3 \epsilon_{\phi} -\epsilon_{Q_B} -\epsilon_{Q_E}-\epsilon_{\chi})\simeq 2(\eta_{\phi} - 3\epsilon_{\phi} - \epsilon_{Q_B}) \,.
\end{eqnarray}
Here $\eta_{\phi}=U_{\phi,\phi}/3 H^2$ and we assume that $\epsilon_{Q_B} \gg \epsilon_{Q_E}, \epsilon_{\chi}$.
One can keep track on the evolution of $\epsilon_H$ by requesting that $\epsilon_{\phi}$ is the dominant  in (\ref{epsH}). 
However,  it is shown  that this condition restricts  significantly the allowed range for $\epsilon_{Q_B}$ \cite{Fujita}.  Instead, Ref. \cite{Fujita} requested 
$\epsilon_{Q_B} <0.2$ given the fact that the central value for $n_s$ measured  by Planck \cite{Planck_cosmo} is $1- n_s \simeq 0.04$. \\

When studying the dynamics of the gauge field,  it is convenient to use the time-dependent mass parameter
of the gauge field fluctuations $m_Q(t)$ and the effective coupling strength $\zeta(t)$ defined as \cite{Emanuela19,Komatsu_new}:
\begin{eqnarray}
\label{mq_zeta}
m_Q (t) \equiv \frac{gQ(t)}{H}\,,\hspace{1cm}
 \zeta(t) \equiv -\frac {\lambda \dot{\chi}(t)} {2 H f} \,,
\end{eqnarray}
that in the slow-roll approximation ($\dot{H} \ll H^2$, $\ddot{\chi} \ll H \dot{\chi}$, $\ddot{Q} \ll H\dot{Q}$) leads to:
\begin{eqnarray}
\label{SR_eq}
 \sqrt{    \frac{ \epsilon_{Q_E} }{ \epsilon_{Q_B} }} \simeq m^{-1}_Q\,, 
 \hspace{0.5cm}  \zeta(t) \simeq m_Q(t)+m_Q^{-1}(t) \,.
\end{eqnarray}

The gauge field fluctuations around the configuration given by Eq. (\ref{gauge_conf}) gives scalar, vector and tensor 
perturbations  \cite{Malek_a,Malek_b}. 
In particular, the tensor perturbations of the gauge field are amplified near the horizon crossing, 
leading to chiral GW background with left- or right-hand sourced tensor modes
\cite{Adshead13,Emanuela19,Thorne,Campetti24}.  
Assuming that only left-hand modes are produced, Ref. \cite{Emanuela19}  shown that the power spectrum of the 
sourced GW tensor modes in the super-horizon limit reads:
\begin{eqnarray}
\label{Ps_sourced}
{\cal P}^{(s)}_t (k) = \frac{\epsilon_{Q_B} H^2}{\pi^2} {\cal F}^2(m_Q)\,,
\end{eqnarray}
where ${\cal F}(m_Q)$ is a monotonically increasing function of $m_Q$ that, using the slow-roll equations (\ref{SR_eq}), can be  approximated by:\\
 ${\cal F}(m_Q) \simeq \exp{ [2.4308m_Q - 0.0218m^2_Q -0.0064m^3_Q-0.86]}\,\,$ for $3 \le m_Q \le 7$. \\
 The tensor-to scalar ratio $r^{(s)}_{k_p}$ of the sourced tensor modes at the peak scale $k_p$  is then:
\begin{eqnarray}
\label{rst}
r^{(s)}_{k_p} & = & \frac{{\cal P}^{(s)}_t}{ {\cal P}^{(v)}_{\zeta}} (k_p) = \frac{\epsilon_{Q_B}H^2}{\pi^2 {\cal P}^{(v)}_{\zeta}}{\cal F}^2(m_Q)\,, 
\end{eqnarray}
where $ {\cal P}^{(v)}_{\zeta}(k_p)$ is the power spectrum of  vacuum curvature fluctuations. 
As ${\cal P}^{(v)}_{\zeta}$ receives negligible
sourced contributions for $m_Q \ge \sqrt{2}$  \cite{Emanuela13,Emanuela19,Komatsu_ref} it can be assumed to be equal to the observed  curvature power spectrum 
${\cal P}^{obs}_{\zeta}$.\\
The tensor-to-scalar ratio $r$ in the model is then given by:
\begin{eqnarray}
\label{r_tot}
r=\frac{ {\cal P}^{(v)}_t + {\cal P}^{(s)}_t} {{\cal P}^{(v)}_{\zeta}}=
\frac{2 g^2 \epsilon_{Q_B}} {\pi^2 m^4_Q  {\cal P}^{v}_{\zeta}} (1+R_{GW}) \,, 
\hspace{1cm} R_{GW}\equiv \frac{ {\cal P}^{(s)}_t} { {\cal P}^{(v)}_t}
= \frac{\epsilon_{Q_B}}{2} {\cal F}^2 (m_Q)
\end{eqnarray},

On the other hand, the shape of the sourced tensor power spectrum depends on the type of axion spectator potential \cite{Fujita_temp}.
For axion potential given in Eq. (\ref{V_axion}) the primordial power spectrum of the
sourced tensor modes, assuming that only left-handed gravitational waves are amplified, has
log-normal shape 
\cite{Thorne,Litebird_new}:
\begin{equation}
\label{Ps_template}
{\cal P}^{(s)}_t (k)=r_*{\cal P}^{(v)}_{\zeta}(k_p) \exp\left[- \frac{1}{2 \sigma^2} \ln^2 \left( \frac{k}{k_p}\right) \right]\,,
\end{equation}
where $r_*$ is the effective tensor-to-scalar ratio at $k_p $ and $\sigma$ is the width of the  bump of sourced tensor power spectrum:
\begin{eqnarray}
\label{rst_sigma}
r_*  = \frac { {\cal P}^{(s)}_t} { {\cal P}^{(v)}_{\zeta}} (k_p)=
\frac{m^4_*H^2}{\pi^2 g^2 {\cal P}^{(v)}_{\zeta}(k_p)}{\cal F}^2(m_*)\,, 
\hspace{1cm} \sigma^2=\frac{ \Delta N^2 }  {2 {\cal G}(m_*)} \,.
\end{eqnarray}
Here $m_* $ represents stable solutions of $m_Q(t)$ equation of motion at $\chi_*=0.5 \pi f$, $\Delta N=\lambda/2 \zeta_*$ and 
${\cal G}(m_*) \simeq 0.666 + 0.81m_*- 0.0145 m^2_*-0.0064 m^3_*$ 
\cite{Emanuela19,Fujita_temp,Thorne}. \\
The power spectrum of sourced GW given in Eq. (\ref{Ps_template}) represents a general prediction of spectator axion-gauge field models if the axion potential has a single inflection point during inflation \cite{Fujita_temp}. 
The validity of this relation in the slow-roll approximation has been checked  in Ref. \cite{Thorne} by comparing with the full numerical solutions of the axion and gauge field background and perturbation equations.  
Eq. (\ref{Ps_template}) uniquely relates  $\{k_p, r_*,\sigma \}$ to the model parameters $\{g,\lambda,\mu, f \}$ once $m_*$ is specified. 

Analytical stable slow-roll solutions for $m_Q$ in absence of backreatiaction  are found in Ref. \cite{Komatsu_new} by solving the equation of motion for $m_Q(t)$.
These solutions are separated in the following  two distinct cases:
\begin{eqnarray} 
\label{Sol}
\hspace{1.5cm}m^A_Q (t)& \simeq & \left[\frac{\kappa\beta(\chi(t))} {3} \right]^{1/3}\,,  
\hspace{3cm} \kappa  \ll 1\,,\\
\hspace{1.5cm}m^B_Q(t) & \simeq & \frac{1}{12} \left[\beta(\chi(t)) + \sqrt{\beta^2(\chi(t)) -144} \right]\,, 
\hspace{0.5cm} \kappa \gg   1\,,  
\end{eqnarray}
where:
\begin{eqnarray}
\label{kapa}
 \beta(\chi(t)) = -\frac{\lambda V_{\chi}(\chi(t))}{H^2 f }\,, \hspace{1cm} \kappa= \left(\frac{gf}{\lambda H}\right)^2\,.
 \end{eqnarray}
 The maximum value $m_*=m_Q(\chi_*)$  is then obtained at  $\chi_*= 0.5 \pi f$ where $\beta_*= \lambda \mu^4/f^2H^2$.

 \subsection{Consistency and backreaction constraints}
 
Several constraints on the axion-SU(2) spectator model has been studied in a number
of papers  \cite{Emanuela19,Komatsu_new,Komatsu_ref,Fujita,Emanuela13,Fujita_back,Malek_SU2,Miz} to ensure that the model is consistent with the viable parameter space for large sourced GW on the
scales constrained by the CMB observations. 

\begin{figure}
\centering
\includegraphics[width=8cm,height=8.cm]{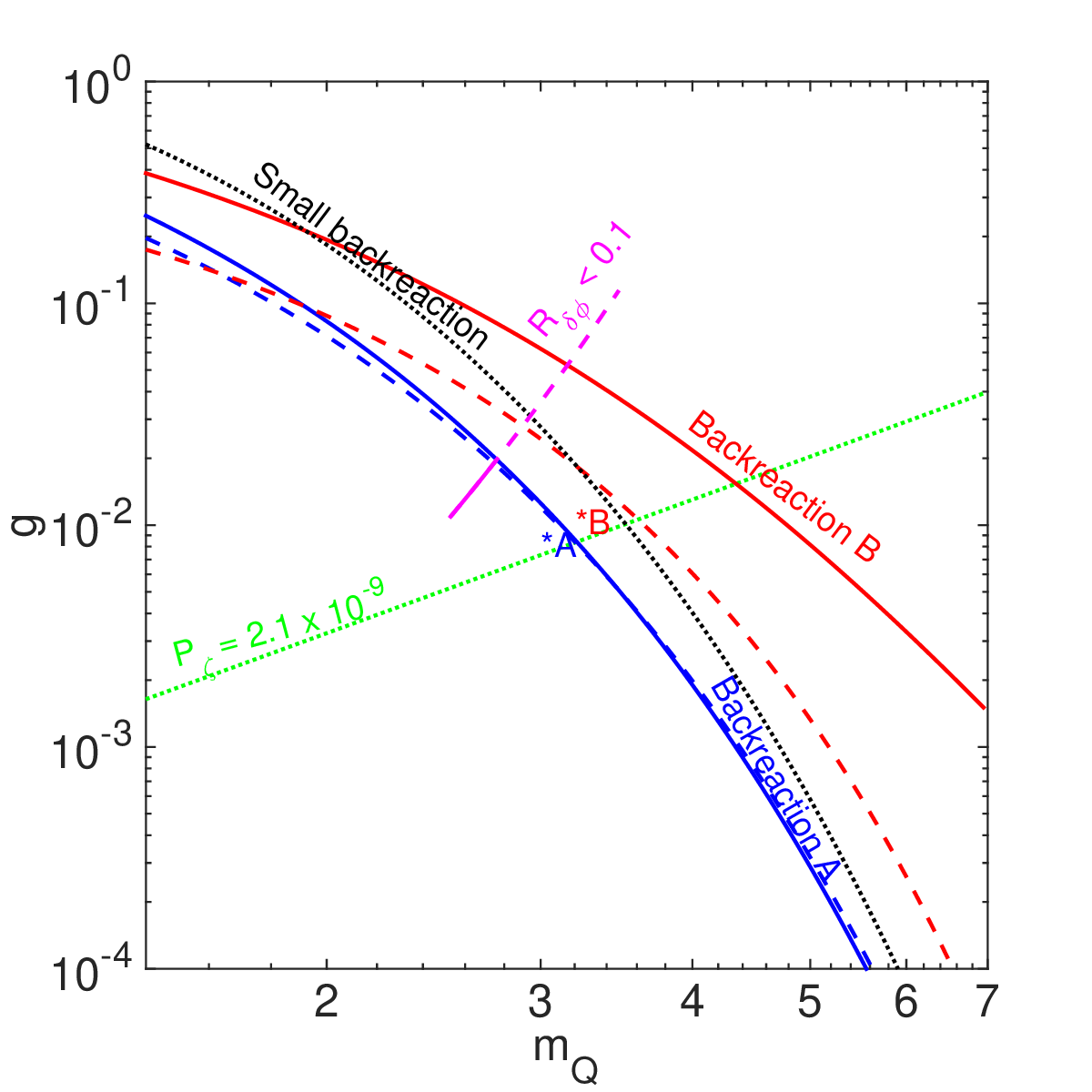}
\caption{Consistency and backreaction constraints:  Lower bound on $g$ from the requirement for
${\cal P}^{(v)}_{\zeta}$ to coincide with ${\cal P}^{obs}_{\zeta}$ (green line).
Upper bound on $g$ from the requirement that the quantum loop corrections to the adiabatic
curvature perturbations ${\cal R}_{\delta \phi}< 0.1$ for $\epsilon_{QB}/\epsilon_{QE}<1$  (magenta continuous line ) 
and $\epsilon_{QB}/\epsilon_{QE}>1$ (magenta dashed line) with $r^{(v})=3.44 \times 10^{-3}$ and $N_k=10$.
Upper bound on $g$ required by the small backreaction regime (black dotted line).
Upper bounds on $g$ from the full numerical computation including  backreaction \cite{Komatsu_new} for: 
{\it i)} the stable solution $m^A_*$ for $\lambda=100$ (continuous blue line) and  $\lambda=50$ (dashed blue line)
and  {\it ii)} the stable solution $m^B_*$ for $\lambda=100$ (continuous red line) and $\lambda=50$ (dashed red line). 
The best fit solutions $m^A_*$ (blue star) and $m^B_*$ (red star) are also indicated.
\label{Fig3} }
\end{figure}

Assuming that $\epsilon_{\chi}$ and $\epsilon_{Q_E}$ are subdominant in $\epsilon_H$, 
 the power spectrum of the curvature perturbations ${\cal P}^{v}_{\zeta}$  can be estimated as  \cite{Komatsu_ref}:
\begin{eqnarray}
\label{Pzeta}
{\cal P}^{v}_{\zeta} \simeq \frac{g^2}{8 \pi^2 m^4_Q}
 \frac{ \epsilon_{\phi} \epsilon_{Q_B}} {(\epsilon_{\phi}+\epsilon_{Q_B})^2}\,.
\end{eqnarray}
The requirement for ${\cal P}^{v}_{\zeta}$ to coincide with ${\cal P_{\zeta}}^{obs}$ leads to the lower bound on $g$:
\begin{equation}
\label{Pzeta}
g \ge g^{min}=\sqrt{32 \pi^2 {\cal P}^{obs}_{\zeta}} m_Q^2\,.
\end{equation}
This constraint is plotted in Figure \ref{Fig3} as function of $m_Q$ with  a  green line.\\

An upper bound on $g$  is coming from the requirement that the quantum  loop corrections 
to the adiabatic curvature perturbations ${\cal R}_{\delta \phi}$ are smaller than the vacuum ones, where  ${\cal R}_{\delta \phi}$ 
approximated as \cite{Komatsu_ref,Komatsu_ref1}:
\begin{eqnarray}
\label{delta_R}
{\cal R}_{ \delta \phi} \simeq \frac{5 \times 10^{-12} }
{ (1+\epsilon_{Q_B} / \epsilon_{\phi} )^2}
e^{7m_Q} m_Q^{11} N^2_k  r^{(v)2}   \,,
\end{eqnarray}
Here $\epsilon_{Q_B}/\epsilon_{\phi}$ is obtain from Eq. (\ref{Pzeta}) , $r^{(v)}$ is the  tensor-to-scalar ratio of the vacuum fluctuations 
and $N_k$ is the number of e-folds during which the axion field is rolling down its potential. 
Depending if $\epsilon_{Q_B}/\epsilon_{\phi}<1$ or $\epsilon_{Q_B}/\epsilon_{\phi}>1$ and demanding  ${\cal R}_{ \delta \phi} <0.1$  
one can obtain a lower bound on $g$ as function of $m_Q$. This constraint is valid for $ 2.5 \le m_{Q} \le 3.3$.\\
Taking $r^{v}=3.44 \times 10^{-3}$ as given by the Higgs-singlet inflation model and $N_k=10$ we obtain 
bounds on $g$ for $\epsilon_{Q_B}/\epsilon_{\phi}<1$ and $\epsilon_{Q_B}/\epsilon_{\phi}>1$, that are plotted in Figure \ref{Fig3} 
with a magenta  lines.

On the other hand, the enhanced  sourced tensor modes result in backreaction terms in the background equations
of motion for axion and gauge fields due to the energy transfer of spin-2 particles produced during inflation
\cite{Emanuela19,Fujita_back,Malek_SU2}. 

The analitycal calculations in the small backreaction approximation \cite{Adshead13} show that this  regime
can be achieved with the following constraint \cite{Emanuela19,Komatsu_ref}:
\begin{eqnarray}
\label{back_small}
g \ll   \left( \frac{24 \pi^2}{2.3 e^{3.9m_Q}} \frac{1}{1+m_Q^{-2}} \right)^{1/2} \,.
\end{eqnarray}
This  constraint is denoted by ``Small backreaction" in Figure \ref{Fig3}. 

More stringent upper bounds of $g$ are numerically obtained in Ref.~\cite{Komatsu_new} 
by solving the background equations of motion for the axion and gauge fields including the backreaction terms.\\
Using the analytic formula for $g^{max}$ given in Ref. \cite{Komatsu_new} we obtained upper bounds on $g$ corresponding to the 
stable solutions  $m^A_*$ and $m^B_*$ for  the gauge field coupling constant  $\lambda=50$\,, $100$. \\
These constraints are denoted in Figure~\ref{Fig3} by `Backreaction A" and ``Backreaction B".

\section{Gravitational waves sourced by the axion-gauge field and Higgs portal interactions}
\label{GW}

 In this section we provide constraints on the parameter space of the spectator axion-gauge field model  
 in presence of Higgs portal interactions. 
 For this purpose we take the Hubble expansion rate $H_{inf}$ during inflation at ${\phi_*} = 5.26M_{pl}$, corresponding to the Hubble crossing of the largest observable CMB scale at $N \simeq 55$ e-folds before the end of inflation. 
 
\begin{figure}
\centering
\includegraphics[width=8cm,height=8.cm]{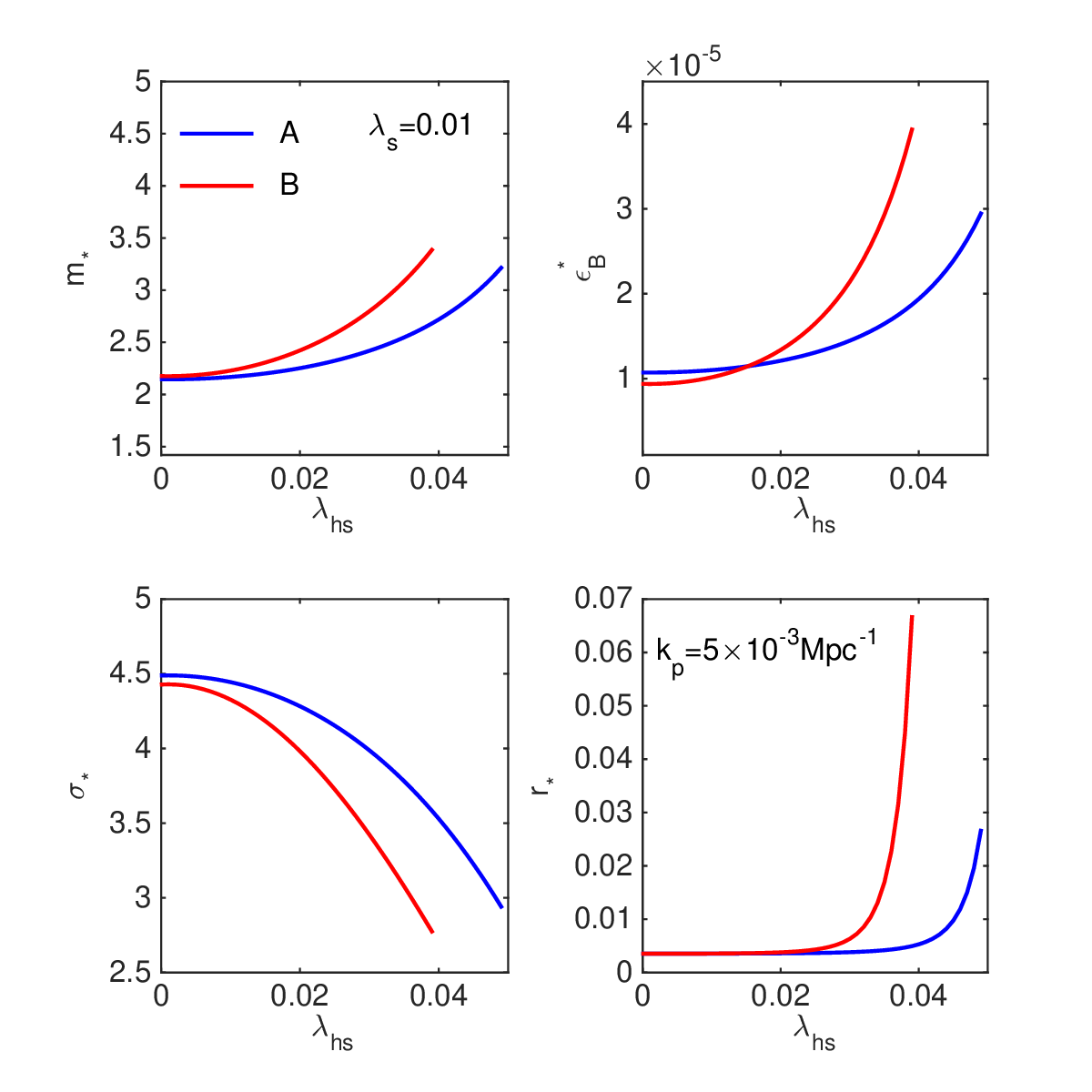}
\caption{  The evolution of  $\{m_*\,,\epsilon^*_B\,,\sigma\,,r_*\}$  with the Higgs portal quartic coupling $\lambda_{hs}$
obtained for the stable solutions of $m_Q$ given in Eq. (\ref{Sol}) 
at their maximum values $m^A_*$ (blue) and $m^B_*$ (red). The sourced tensor-to-scalar ratio $r_*$ is obtained at $k_p=5 \times 10^{-3}$Mpc$^{-1}$.
We take $\lambda_s=0.01$ and fix the axion-gauge field model parameters to:
$g\,,\lambda\,,\mu\,,f\,=\{10^{-2}\,,50\,, 4 \times10^{-4}\,,10^{-2} $\}. 
The Hubble expansion rate during inflation is fixed at $ \phi_* = 5.26 M_{pl}$
corresponding to the Hubble crossing of the largest observable CMB scale at $N \simeq 55$ e-folds before the end of inflation. 
\label{Fig4}}
\end{figure}

We evaluate the impact of 
Higgs portal interaction on parameters of the GW sourced tensor modes power spectrum
$\{m_*\,,\epsilon^*_B\,,\sigma\,,r_*\}$ obtained for the stable solutions $m_Q$ given in Eq. (\ref{Sol}) 
at their maximum values $m^A_*$ and $m^B_*$. \\
Figure \ref{Fig4} presents the evolution of these parameters with the Higgs-singlet quartic coupling $\lambda_{hs}$ for $\lambda_s=0.01$ and 
the spectator axion-gauge field model parameters $g\,,\lambda\,,\mu\,,f\,=\{10^{-2}\,,50\,, 4 \times10^{-4}\,,10^{-2}\}$.
All  parameters show dependences on $\lambda_{hs}$ that are enhanced for the $m^B_*$ solution, with clear impact for the values of
the sourced tensor-to-scalar ratio $r_*$. 
The figure shows that the Higgs portal interactions could enhance the GW  signal sourced by the gauge field fluctuations 
in the CMB B-mode power spectra.

We perform an inference of the model parameters using the  Monte-Carlo Markov Chains (MCMC) approach, accounting for
consistency and backreaction constraints discussed in the previous section.\\
To obtain the variation intervals  of the axion-gauge field parameters required by the MCMC analysis
we use the confidence intervals of the Higgs-singlet inflation model given in Eq. (\ref{confidence}). \\
The  variation interval for the gauge coupling $g$ is obtained as follows:
$g_{min}$ is calculated at $m_*=2.5$ from the requirement that ${\cal P}_{\zeta}$ should coincide with ${\cal P}^{obs}_{\zeta}$ 
as given by Eq. (\ref{Pzeta}), 
while $g^{max}$ is obtained at $m_*=3.5$ from the requirement  ${\cal R}_{\delta \phi} \le 0.1$ using Eq. (\ref{delta_R}), as presented in Section~2.3.\\
The  variation interval for the axion decay constant $f$ is  obtained from the requirement  $\kappa \lessgtr 1$,
 where $\kappa$ is given by Eq.~(\ref{kapa}), 
for  $\lambda_{min}=30$ and $\lambda_{max}=100$. \\
The variation intervals for the modulation amplitude of the axion potential $\mu$ are calculated using the extrema values of
 $H_{inf}$ given in Eq. (\ref{confidence}) and those  of $g$, $f$, $\lambda$ and $m_*$ discussed above.
 We obtain these bounds for  both $m^A_Q$ and $m^B_Q$ solutions. 
The intervals for axion-gauge field model parameters used in this analysis are given below:
\begin{eqnarray}
\label{intervals}
g &:& \hspace{0.5cm}1.07\times 10^{-3} \,\,\, \div \,\,\,1.98\times 10^{-2}  \\
f &: & \hspace{0.5cm} 9.81\times 10^{-3}\,\,\, \div \,\,\,3.33 \times 10^{-1} \nonumber \\
\mu^{A} & : &  \hspace{0.5cm}9.61 \times 10^{-5}M_{pl} \,\,\, \div \,\,\, 2.33 \times 10^{-3}M_{pl} \nonumber\\
\mu^{B} & : &  \hspace{0.5cm}1.14 \times 10^{-4}M_{pl} \,\,\, \div \,\,\, 1.14 \times 10^{-3}M_{pl} \nonumber \\
\lambda & :  &\hspace{0.5cm} 30\,\,\, \div  \,\,\,100 \nonumber
\end{eqnarray}

We modify the Boltzmann CAMB code \cite{camb} to compute $r_*$ and $\sigma$ given in Eq.(\ref{rst_sigma}) and to evaluate the sourced GW  
power spectra from Eq. (\ref{Ps_template}) for the stable solutions $m^A_*$ and $m^B_*$, in the parameter 
intervals indicated  in Eqs.(\ref{confidence}) and  (\ref{intervals}). 
Our goal is to infer  the axion-gauge field model parameter space for both solutions and to
evaluate their impact on the CMB B-mode polarization power spectra. 

\begin{figure}
\begin{subfigure}[b]{0.5\textwidth}
\centering

\vspace{-3.5cm}
\includegraphics[width=6.cm,height=6.cm]{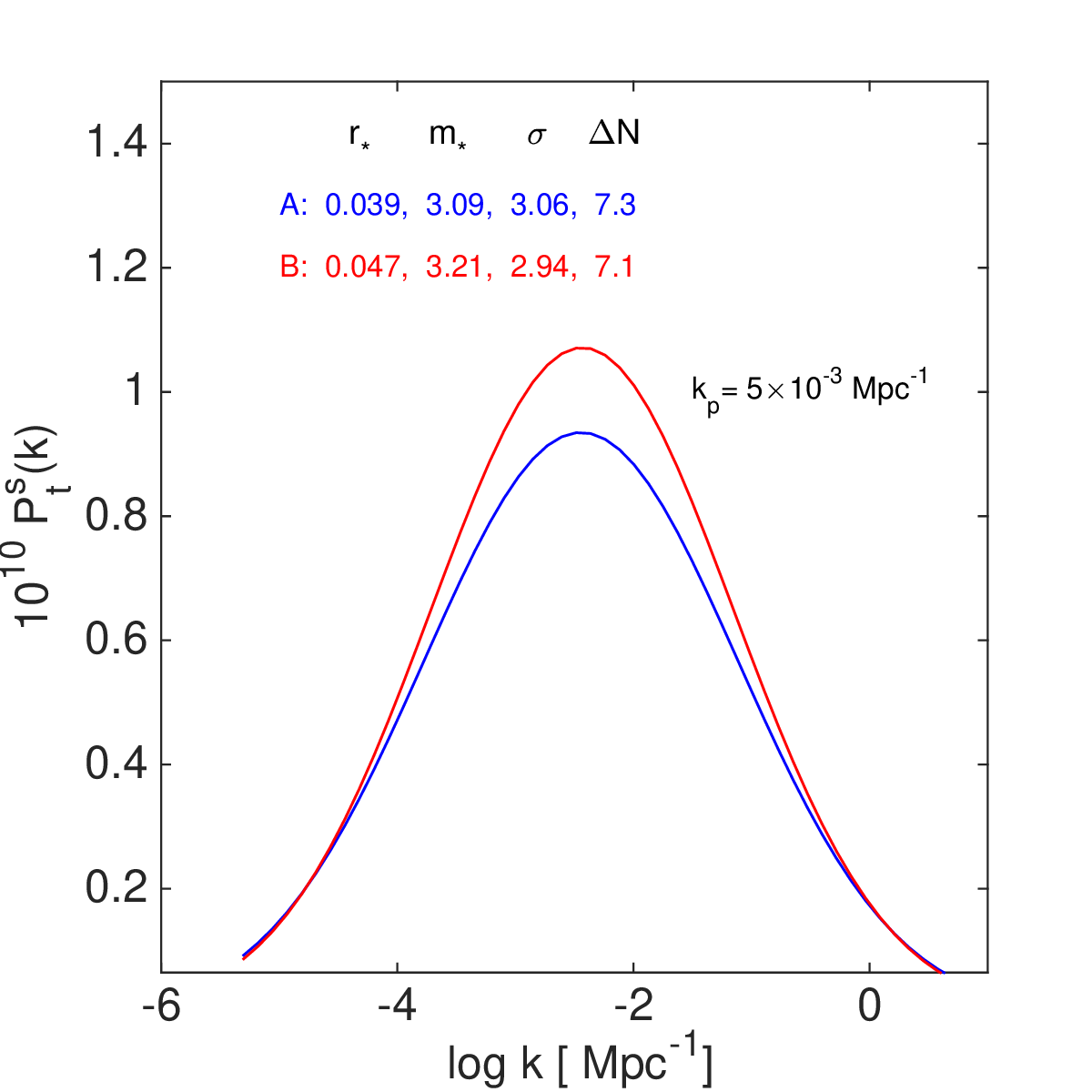}
\end{subfigure}
\begin{subfigure}[b]{0.55\textwidth}
\centering

\vspace{-3.5cm}
\includegraphics[width=6.cm,height=6.cm]{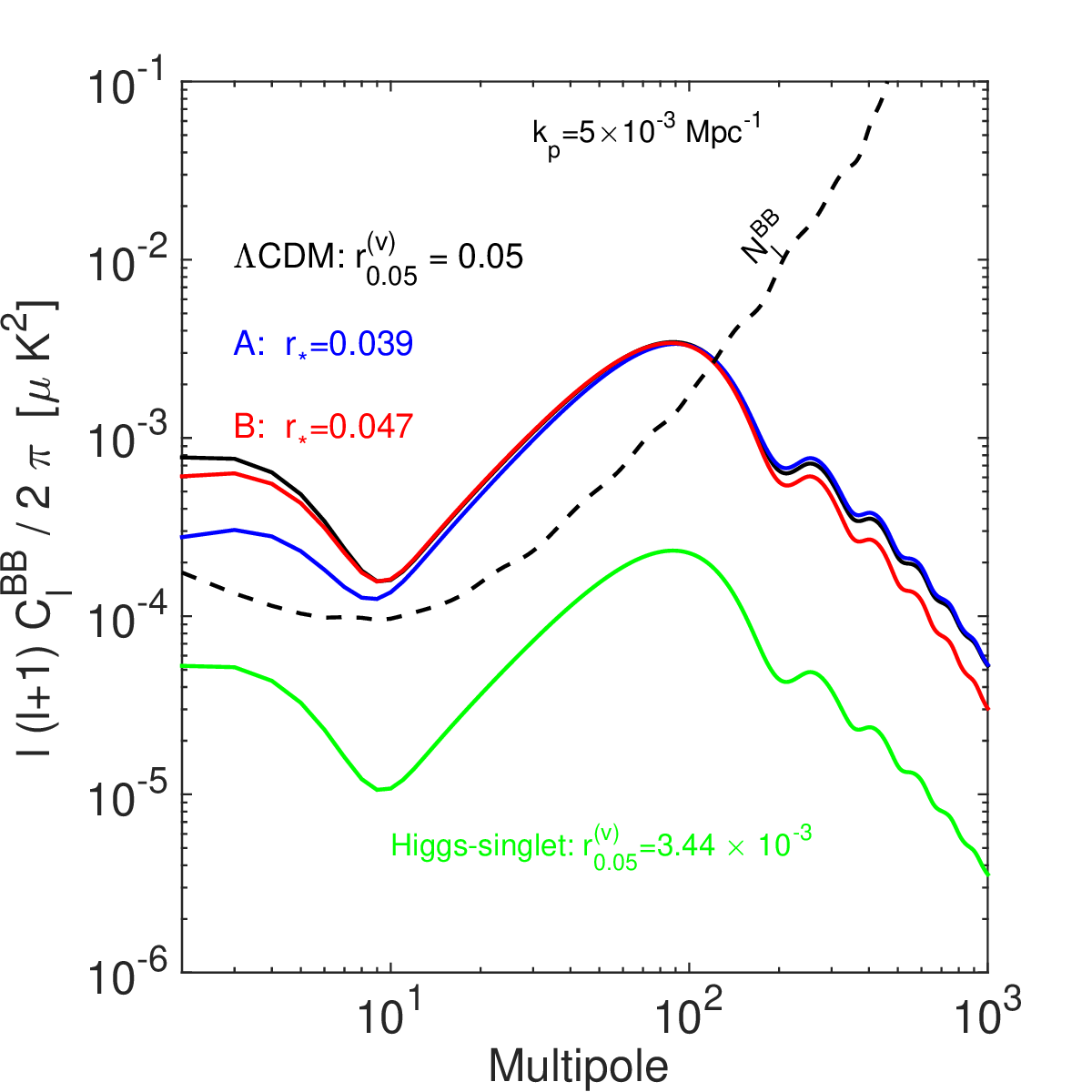}
 \end{subfigure}
\caption{ {\it Left:} The best fit of sourced GW power spectra ${\cal P}^{(s)}_t $          
for $m^A_*$ (blue line) and $m^B_*$ (red line) solutions. 
The best fit  parameters of ${\cal P} ^{(s)}_t$ are also indicated. 
{\it Right:} The corresponding  best fit  B-mode polarization power spectra  for $m^A_*$ (blue line) and $m^B_*$ (red line) solutions. 
For comparison, the B-mode polarization power spectra of $\Lambda$CDM target model (black line) and  Higgs-singlet inflation model (green line) are presented.The noise power spectrum $N^{BB}_l$ (black dashed line) corresponding to the LiteBird mission experimental configuration \cite{Thorne} is also 
shown.  \label{Fig5}. }   
\end{figure}

\begin{figure}
\begin{subfigure}[b]{0.28\textwidth}
\centering

\includegraphics[width=8.cm,height=4.cm]{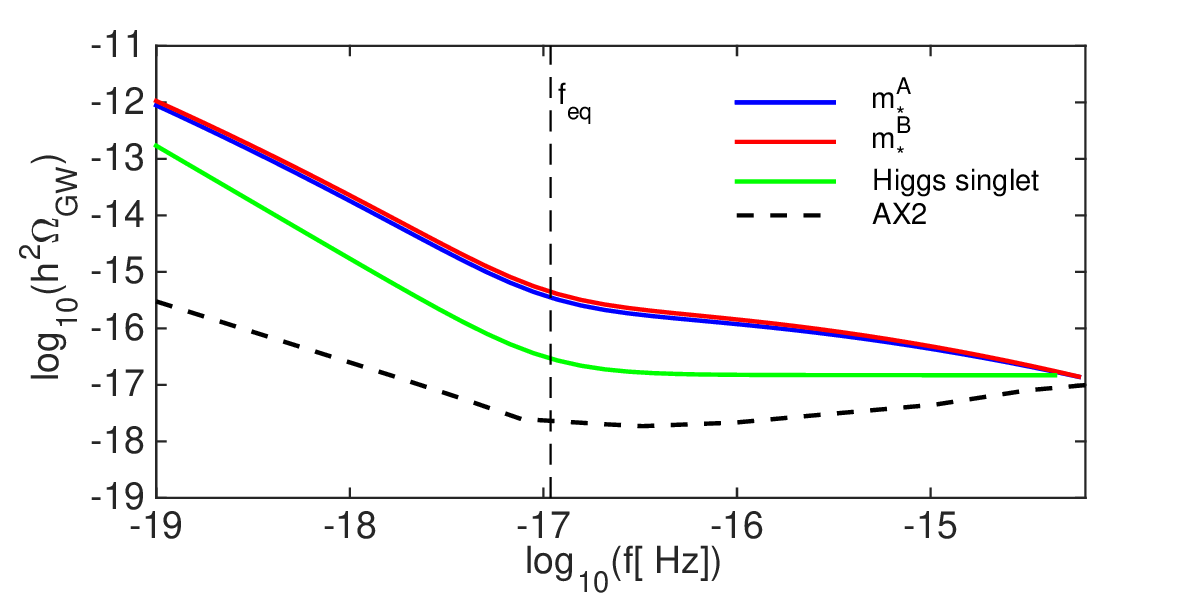}
\end{subfigure}
\begin{subfigure}[b]{1.\textwidth}
\centering

\includegraphics[width=7.cm,height=4.cm]{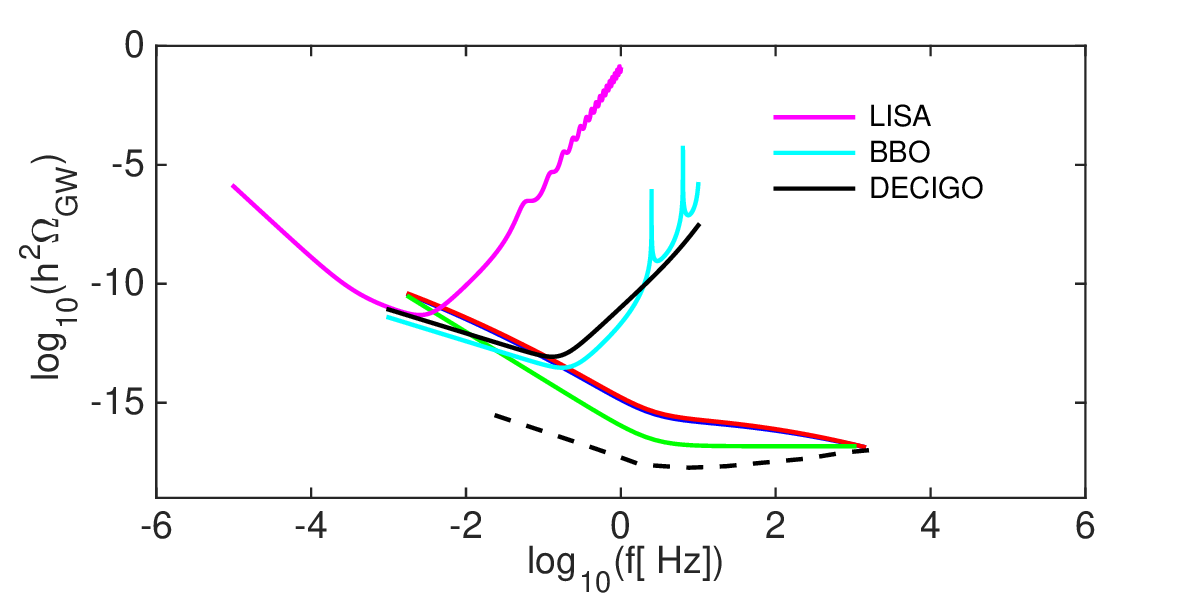}
 \end{subfigure}
\caption{ {\it Left}: Evolution with frequency of the  energy density parameter $h^2\Omega_{GW}(f)$  
of the sourced primordial GW for $m^A_*$ (blue line)  and $m^B_*$ (red line) best fit solutions obtained 
for the LiteBird observing strategy. 
The solid green line shows the vacuum energy contribution of Higgs-singlet model. 
In all cases the GW energy density spectra at present time are adiabatic 
with a slope that change at frequency scales that make the transition between matter and radiation domination eras.  
For comparison we also show $h^2\Omega_{GW}(f)$ sourced by axion-SU(2) gauge field model AX2
from  Ref. \cite{Daniela}. 
 {\it Right}: The $h^2\Omega_{GW}(f)$ power spectra presented in the left panel,  rescaled at GW frequencies corresponding 
 to 15 e-folds before the end of inflation, and the sensitivity curves for 
 LISA, DECIGO, and BBO interferometers. \label{Fig6}}   
\end{figure}

To address the detectability of the GW sourced by the gauge field in presence of Higgs portal interactions we take as target model the {\sc Planck} best fit 
$\Lambda$CDM model \cite{Planck_cosmo} with the vacuum tensor-to-scalar ratio $r^{(v)}=0.05$ at $k_0=0.05$Mpc$^{-1}$
and the normalisation $P^{obs}_{\zeta}=2.1 \times 10^{9}$. We also take the noise power spectrum for the experimental configuration of  the LiteBird mission given in Ref. \cite{Thorne}. \\
As the sourced tensor modes are expected to exceed  the vacuum contribution at large CMB observable scales, 
 we take in this analysis  the B-modes polarization power spectra in the multipole interval ${\it l} \simeq (2\,\, \div \,\,150)$ and evaluate 
the GW sourced tensor-to-scalar ratio at $k_p=5 \times 10^{-3}$.
As before, we use the Monte-Carlo Markov Chains (MCMC) technique to sample from the space of axion-gauge field and Higgs portal parameters
and generate estimates of their posterior distributions. As mentioned, we assume a flat universe and uniform priors for all parameters adopted in the analysis. \\
Left panel from Figure \ref{Fig5} presents the best fit sourced GW power spectra for  $m^A_*$ and $m^B_*$ solutions, while 
right panel  from the same figure shows the corresponding  B-mode polarization power spectra.  
For comparison, the B-mode polarization power spectra of $\Lambda$CDM target model and  Higgs-singlet inflation model are also presented. \\
The spectrum of the sourced GW energy density  at the present time and 
at a given frequency $f=k/2\pi$ can be approximated as  \cite{Caprini}:   
\begin{eqnarray} 
\label{omega_gw}
h^2\Omega_{GW}(f) =  \frac{3}{128} \Omega_{rad} {\cal P}^{(s)}_t(f)
 \left[ \frac{1}{2}   \left( \frac{f_{eq}}{f} \right )^2 +\frac{16}{9} \right ] \,,
\end{eqnarray}        
where $h^2$ is defined such that  $ H_0 =100 h$ km s$^{-1}$ Mpc$^{-1}$
is the Hubble parameter at the present time, ${\cal P}^{(s)}_t(f)$ is power spectrum of the sourced tensor modes,
$\Omega_{rad}=(1 - f_{\nu})^{-1} \Omega_{\gamma} $ is the present radiation energy density parameter, 
$(1-f_{\nu})^{-1} = 1.68$ for the SM expectation value of $N_{eff}=3.046$ relativistic degrees of freedom \cite{Salas},  
$\Omega_{\gamma}= 2.38 \times 10^{-5} h^{-2}$ is the present photon energy density parameter
and 
$f_{eq}=1.09 \times 10^{-17}$ Hz is the frequency entering the horizon at matter-radiation equality. 
We use $f/Hz = 1.5 \times 10^{-15} k/Mpc^{-1}$ .  \\
Left panel from  Figure \ref{Fig6} presents the evolution with frequency of the  energy density parameter $h^2\Omega_{GW}(f)$ of the sourced primordial GW for $m^A_*$  and $m^B_*$  best fit solutions obtained 
for the LiteBird observing strategy.
The solid green line shows the vacuum energy contribution of Higgs-singlet model. 
In all cases the GW energy density spectrum at present time is adiabatic 
with a slope that change at frequency scales that make the transition between matter and radiation domination eras.  \\
For comparison we also show $h^2\Omega_{GW}(f)$ sourced by axion-SU(2) gauge field model AX2
from  Ref. \cite{Daniela}. For all cases the GW energy spectra are adiabatic with a slope that change at frequency scales 
corresponding to modes entering the horizon during the matter-radiation equality.\\
In the right panel from Figure \ref{Fig6}  we show the same  $h^2\Omega_{GW}(f)$ power spectra along with 
the sensitivity curves of the future  satellite-borne GW interferometers LISA \cite{Lisa}, DECIGO \cite{DECIGO} 
and BBO \cite{BBO}. As $h^2\Omega_{GW}(f)$ power spectra covers 55 e-folds before the end of inflation,
to provide a comparison with the sensitivity of LISA, DECIGO and BBO, we rescale the GW frequency to 15 e-folds 
before the end of inflation \cite{Oxana}.\\
The  sensitivity curves of the GW interferometers are obtained by using the `strain noise power spectra" file available online
 in Zenodo repository \cite{Fresh1,Zendo}. The $h^2\Omega_{GW}(f)$ power spectra obtained for $m^A_*$ and $m^B_*$ best fit solutions obtained for the LiteBird observing strategy are potentially detectable in the frequency rage $10^{-2}$Hz - 1 Hz.

Table~1 presents the mean values and absolute errors at 68\% confidence of the model parameters obtained from the MCMC analysis
for $m^A_*$ and $m^B_*$ solution.
We find that  the tensor-to-scalar ratio of the sourced GW in presence of Higgs portal interactions is enhanced 
to a level that overcomes the  vacuum tensor-to-scalar ratio by a factor ${\mathcal O}$(10) for both solutions, 
much above the detection threshold of the near-future B-modes polarization LiteBird experiment, 
in agreement with the CMB observations on curvature fluctuations and with the allowed parameter space of Higgs portal interactions.

\begin{figure}
\centering

\vspace{-3.cm}
\includegraphics[width=17cm,height=17.cm]{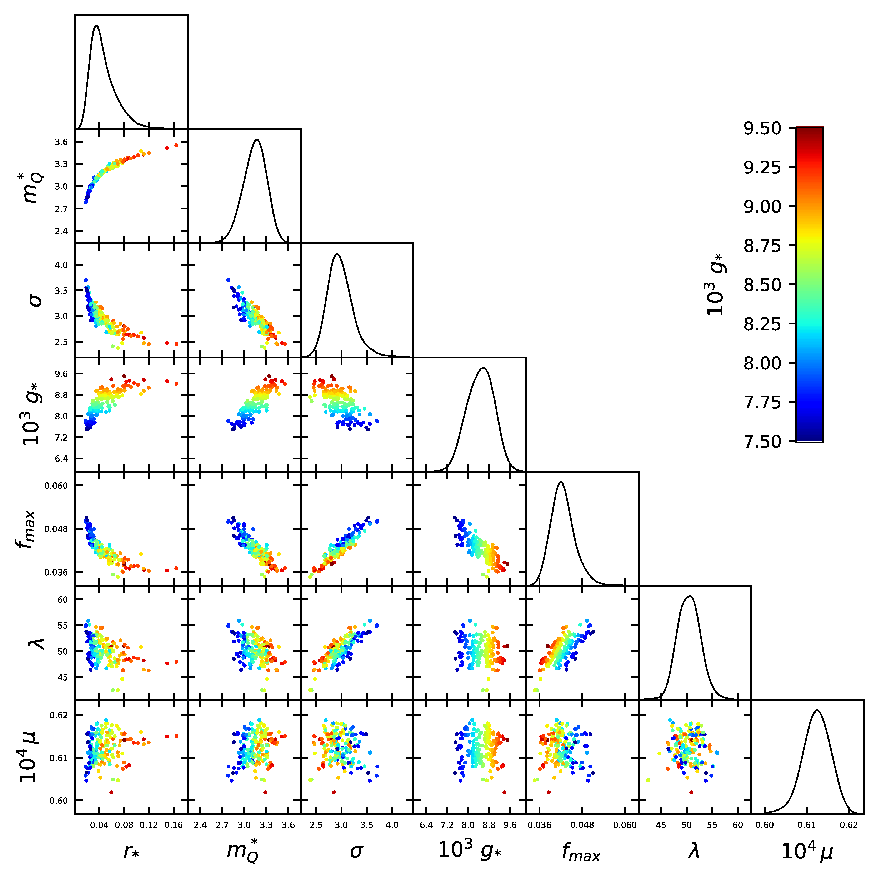}
\caption{ The marginalised probability distributions obtained for the parameters of the axion-gauge field spectator model with Higgs portal interactions
for $m^A_*$ solution at $k_p~=~5~\times~10^{-3}$~Mpc~$^{-1}$. \label{Fig7} }
\end{figure}

\begin{figure}
\centering

\vspace{-5cm}
\includegraphics[width=17cm,height=17.cm]{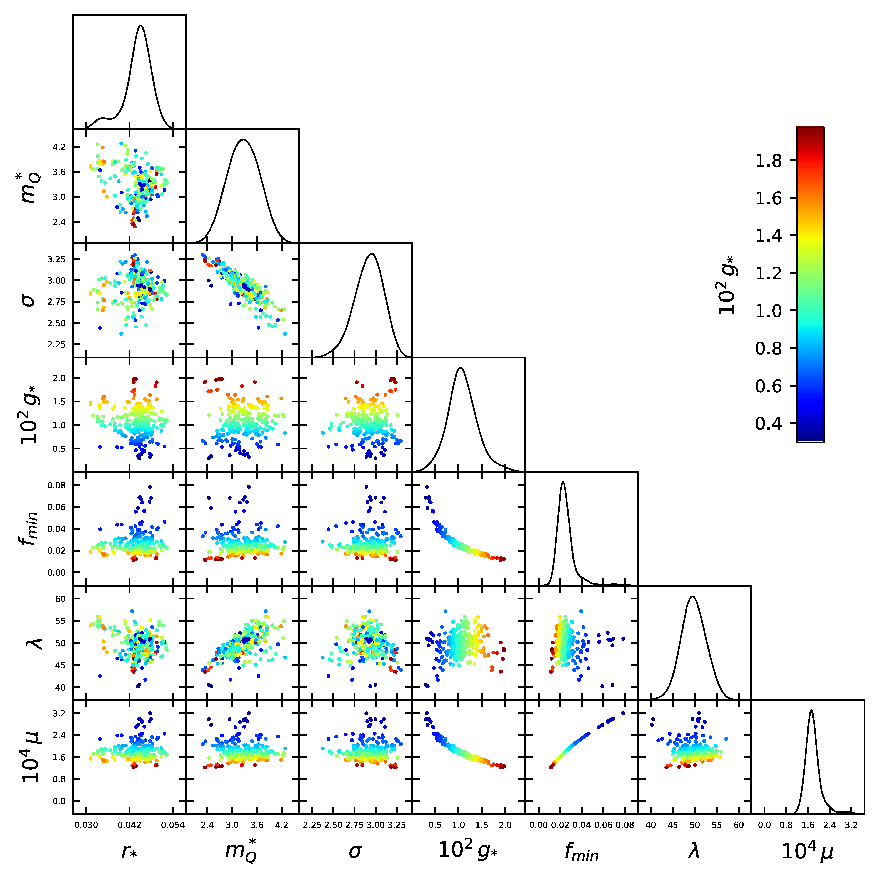}
\caption{ The marginalised probability distributions obtained for the parameters of the axion-gauge field spectator model with Higgs portal interactions
for $m^B_*$ solution at $k_p~=~5~\times~10^{-3}$~Mpc~$^{-1}$. \label{Fig8} }
\end{figure}

In Figure \ref{Fig7} and Figure \ref{Fig8} we present the  marginalised probability distributions of the axion-gauge field spectator 
model with Higgs portal interactions for the $m^A_*$ and $m^B_*$  solutions. The figures show the correlation between axion-gauge field model parameters $\{g,\,\lambda,\, \mu,\, f\}$ and the power spectrum of the sourced tensor modes parameters  $\{r_{*},\, \sigma,\, m_*\}$ at $k_p=5 \times 10^{-3}$Mpc$^{-1}$.

\begin{figure}

\vspace{-2cm}
\centering
\includegraphics[width=8cm,height=8.cm]{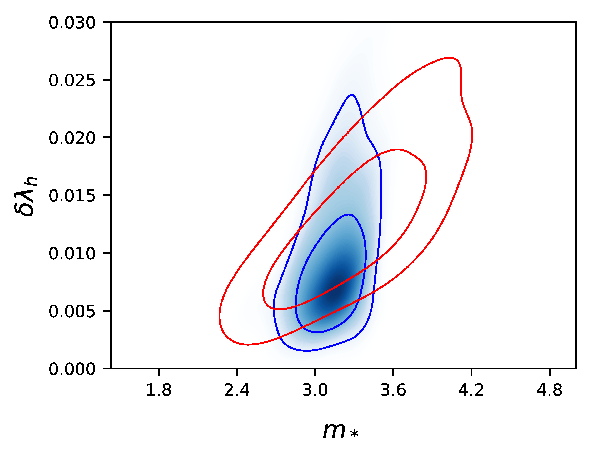}
\caption{ The 2D marginalised probability distributions obtained for
 $m^A_*$ (blue) and $m^B_*$ (red) solutions at  68\% and 95\% confidence intervals. The figure shows the correlations between $m_*$  and the 
threshold correction of the Higgs quartic coupling $\delta \lambda_h =\lambda^2_{hs}/ 4 \lambda_s$.
\label{Fig9}}
\end{figure}

Figure \ref{Fig9} presents the 2D marginalised probability distributions in $m_* ~-~ \delta \lambda_h$ plane for
 $m^A_*$ (blue) and $m^B_*$ (red) solutions 
at  68\% and 95\% confidence intervals. The figure shows the correlations between $m_*$  and the 
threshold correction of the Higgs quartic coupling $\delta \lambda_h =\lambda^2_{hs}/ 4 \lambda_s$.

\begin{table}
\caption{ The mean values and absolute errors of the parameters of the axion-gauge field spectator model with Higgs portal interactions
obtained at  $k_p~=~5~\times~10^{-3}$~Mpc~$^{-1}$ for $m^A_*$ and $m^B_*$  solutions. The errors are quoted at 68\% CL. }
\begin{center}
\begin{tabular}{|l|c|c|}
\hline
 Parameter&$ m^A_*$  & $m^{B}_*$ \\
 \hline
$r_*$&   0.039 $\pm$   0.0027 & 0.047 $\pm$ 0.0031     \\
$m_*$& 3.091  $\pm$  0.035&  3.201 $\pm$  0.036      \\
$\sigma$& 3.061 $\pm $ 0.185 &   2.940 $\pm$0.164      \\
\hline
$g$& $(8.32 \pm 0.048) \times 10^{-3}$& (1.01 $\pm$ 0.03) $\times 10^{-2}$ \\    
$\lambda$&50.85$ \pm$ 1.34&  49.46 $\pm$1.49              \\
$\mu$&(6.21 $\pm$ 0.12) $\times 10^{-4}$ & (2.15 $\pm$ 0.28) $\times 10^{-4}$         \\
$f$&  (4.37 $\pm$ 0.31)$\times 10^{-2}$& (2.15 $\pm$ 0.17) $\times 10^{-2}$ \\
\hline
$\lambda_s$& 0.084 $\pm$ 0.011&0.071 $\pm$ 0.022 				\\
$\lambda_{hs}$ & 0.049 $\pm$ 0.002& 0.051 $\pm$ 0.001	\\
$\delta_{\lambda_h}$&(7.11 $\pm$ 0.021) $\times 10^{-3}$ &(1.18 $\pm$0.03) $\times 10^{-2}$			\\
\hline
\end{tabular}
\end{center}
\end{table}

\section{Conclusions}
\label{final}

In this work we investigate a
 scenario where an axion field and an non-Abelian gauge field are confined to the spectator 
sector while the inflation sector is represented by a mixture of Higgs boson and a scalar singlet with large non-zero {\it vev}
 that are non-minimally coupled to gravity. 
We assume that there is no coupling, up to gravitational interactions, between inflation and spectator sectors, 
the background energy density is dominated by the inflaton and the Hubble expansion is constant during inflation.

We place constraints on Higgs-singlet model parameters from the requirement to satisfy the observational bounds on the curvature perturbations.
We show that a mixture of Higgs boson with a heavy scalar singlet with large {\it vev} is a viable model of inflation 
that satisfy the existing observational data and the perturbativity constraints, avoiding at the same time the EW vacuum metastability
as long as the Higgs portal interactions lead to positive tree-level threshold corrections for SM Higgs quartic coupling. 

We evaluate the impact of the Higgs quartic coupling threshold corrections on the GW sourced tensor modes power spectrum 
for two stable slow-roll solutions  for the mass parameter of gauge field fluctuations  \cite{Komatsu_new}
while accounting for the consistency and backreaction constraints. \\
We show that the Higgs portal interactions enhance the GW  signal sourced by the gauge field fluctuations 
in the CMB B-mode ploarization power spectra. 

To address the detectability of the GW  sourced by the gauge field fluctuations in presence of Higgs portal interactions 
we take as target model the {\sc Planck} best fit 
$\Lambda$CDM cosmology \cite{Planck_infl,Planck_cosmo}  with the vacuum tensor-to-scalar ratio $r^{(v)}=0.05$ at $k_0=0.05$Mpc$^{-1}$ and 
the noise power spectrum for the experimental configuration 
of  the LiteBird mission  \cite{Thorne}. Using the MCMC technique 
we sample from the parameter spaces of Higgs-singlet and axion-gauge field model
and generate estimates of model parameters from their posterior distributions.
We obtain in this way comprehensive  constrains on Higgs portal, axion-gauge field  and sourced  GW power spectra parameters. 

We find that the tensor-to-scalar ratio of the sourced GW in presence of Higgs portal interactions is enhanced to a level that overcomes the vacuum tensor-to-scalar ratio by a factor $\mathcal O$(10), much above the detection threshold of the near-future B-modes polarization LiteBird experiment, in agreement with the CMB observations on curvature fluctuations and with the allowed parameter space of Higgs portal interactions.

A large enhancement of the GW sourced can be also detected 
by experiments such as pulsar timing arrays and laser/atomic interferometers \cite{Thorne,Daniela}.\\
On the other hand, a significant Higgs-singlet mixing can be probed at LHC 
by  the measurement of production cross sections for Higgs-like states \cite{Falk,PDG},
while a significant tree level threshold of the Higgs quartic coupling can
be measured at colliders \cite{ATLAS,CMS}.

\section{Acknowledgment}
 The author acknowledges the
 use of the GRID computing facility at the Institute of Space Science. \\
 This research was supported by the Romanian Ministry of Research,
Innovation and Digitalization under the Romanian National Core Program
LAPLAS VII - contract no. 30N/2023.

\end{document}